\RequirePackage{fix-cm}
\documentclass{svjour3}                     % onecolumn (standard format)
\smartqed
\usepackage[numbers]{natbib}

\usepackage[pdftex]{graphicx}
\usepackage[cmex10]{amsmath}
\usepackage[ruled,vlined,linesnumbered]{algorithm2e}
\usepackage{subcaption}
\usepackage{xspace}
\usepackage{color}
\begin{document}

\title{Towards Finding Non-obvious Papers: An Analysis of Citation Recommender Systems}
\titlerunning{Towards Finding Non-obvious Papers: An Analysis of Citation Recommender Systems}

\author{Haofeng Jia \and  Erik Saule}
\authorrunning{Jia et al.}

\institute{H. Jia \and E. Saule\at
           Dept. Computer Science\\
	   University of North Carolina at Charlotte\\
	   \email{\{hjia1,esaule\}@uncc.edu}
}
%\date{Received: \today  / Accepted: }
\maketitle

\begin{abstract}
As science advances, the academic community has published millions of research papers.
Researchers devote time and effort to search relevant
manuscripts when writing a paper or simply to keep up with current research. In this paper, we consider the problem of citation recommendation by extending
a set of known-to-be-relevant references. Our analysis shows the degrees of cited papers in the subgraph induced by the citations of a paper, called projection graph, follow
a power law distribution. Existing popular methods are only good at finding the long tail
papers, the ones that are highly connected to others. In other words, the majority of cited papers are
loosely connected in the projection graph but they are not going to be found by existing methods. To address this
problem, we propose to combine author, venue and keyword information
to interpret the citation behavior behind those loosely connected papers. Results show that
different methods are finding cited papers with widely different properties. We suggest multiple recommended lists
by different algorithms could satisfy various users for a real citation recommendation system.
Moreover, we also explore the fast local approximation for combined methods in order to improve the efficiency.
\keywords{Citation recommendation \and Graph analysis \and
Academic graph \and Local approximation \and Projection Graph}
\end{abstract}

%%%%%%%%%%%%%%%%%%%%%%%%%%%%%%%%%%%%%%%%%%%%%%%%%%%%%%%%%%%%%%%%%%%%%%%%%
\section{Introduction}
\label{sec:intro}

Scientists around the world have published tens of millions of research papers, and the number of new papers
has been increasing with time. For example, according to DBLP~\cite{DBLP:journals/pvldb/Ley09}, computer scientists published 3 times more
papers in 2010 than in 2000. At the same time, literature search became an essential task performed daily by
thousands of researcher around the world. Finding relevant research works from the gigantic number of published
articles has become a nontrivial problem.\looseness=-1

Currently, many researchers rely on manual methods, such as keyword search via Google Scholar~\footnote{https://scholar.google.com/} or Mendeley~\footnote{https://www.mendeley.com/}, to discover new
research works. However, keyword based searches might not be satisfying for two reasons: firstly, the vocabulary gap
between the query and the relevant document might results in poor performance; secondly, a simple string of keywords
might not be enough to convey the information needs of researchers. There are many instances where such a keyword
query is either over broad, returning many articles that are loosely relevant to what the researcher actual need,
or too narrow, filtering many potentially relevant articles out or returning nothing at all~\cite{el2011beyond}.

To alleviate the above mentioned problems, many research works proposed citation recommendation algorithms which use a manuscript
instead of a set of keywords as query~\cite{strohman2007recommending,he2010context,he2011citation,lu2011recommending,huang2012recommending}. For example, context-aware citation recommendation is designed to recommend relevant papers for
placeholders in the query manuscript based on local contexts~\cite{he2010context,he2011citation}.
Manuscript based citation recommendation is great to help with the writing process. However, we are interested here in helping the research process which usually comes long before manuscripts are fleshed out.
Researchers have devoted efforts on citation recommendation based on a set of seed papers~\cite{mcnee2002recommending,torres2004enhancing,gori2006research,caragea2013can,Kucuktunc13-ASONAM}. Most approaches rely on the citation
graph to recommend relevant papers, such as collaborative filtering~\cite{mcnee2002recommending} and random walk framework~\cite{gori2006research}. The different approaches to recommending academic papers have been extensively surveyed by~\cite{Beel2016}.

We consider in this paper the problem of extending a set of known
references, which is helpful in recommender system and
academic search engine, such as theAdvisor~\cite{Kucuktunc13-JCDL}.
We show that classic methods (namely, PaperRank and Collaborative
Filtering) perform reasonably well, but have an inherent bias. Because
they base their decision on citation patterns, they tend to only find
papers that have many links to the known references, a set of
papers that are obvious. Unfortunately, less than half of the references
of a paper are connected to more than two other references. This causes the
algorithms to ignore lightly connected papers despite being half of the references in practice.\looseness=-1

We propose to use metadata information, such as authorship and textual
information, to identify the non-obvious connections between papers. Besides meta path
based methods, we design two types of algorithms. One type only uses metadata; logAVK directly scores
candidate papers based on the similarity of their metadata to the query papers. And one extends PaperRank with some
metadata; C+X extends PaperRank by adding author, venue or keyword nodes in the graphs to enable
random walk paths between papers with similar metadata. Moreover, various keywords extraction strategies
for C+K are investigated.

Our experiments show that the methods that extend PaperRank can improve
the quality of the recommendation. Also logAVK provides a
different perspective on the queries, despite it does not score as
well in various quality metrics as other algorithms.\looseness=-1
For a practical citation recommendation system, the efficiency
of underlying recommendation algorithm is also important. Therefore,
we propose local C+X methods which is 15x faster while they are as
effective as original C+X methods.

A preliminary version of this work was published in~\cite{Jia17}. We
extend in this paper with following major extensions:
\begin{itemize}
\item We conduct additional experiments on 7 different meta path-based approaches.
\item We investigate the impact of various keywords extraction strategies on C+K;
\item We explore the fast local approximation methods for C+X in order to improve the efficiency.
\end{itemize}

The paper is organized as follows: we introduce
related work in Sec.~\ref{sec:relwork}. In
Sec.~\ref{sec:problemstatement}, we define the citation
recommendation problem and present existing
methods. Based on the analysis of Sec.~\ref{sec:globalperf}, we
propose to use metadata  to enhance the performance on
loosely connected papers in
Sec.~\ref{sec:finding-loosely}. Sec.~\ref{sec:usefulness} argues the use of
the different algorithms in a practical system. Finally,
Sec.~\ref{sec:ccl} discusses the usefulness
in real systems.\looseness=-1
\section{Related Work}
\label{sec:relwork}

Many works addressed citation recommendation.

%\subsection{Citation Recommendation}
\emph{Seed papers based citation recommendation.}
Given a "basket" of citations, McNee et al. \cite{mcnee2002recommending} explore the use of
Collaborative Filtering (CF) to recommend papers that would be suitable additional references
for a target research paper. They create a ratings matrix where citing papers correspond to
users and citations correspond to items. The experiments show CF could generate high quality
recommendations.
As a follow-up, Torres et al. \cite{torres2004enhancing} describe and test different techniques
for combining Collaborative Filtering and Content-Based Filtering.
A user study shows many of CF-CBF hybrid recommender algorithms can generate research paper
recommendations that users were happy to receive. However, offline experiments show those
hybrid algorithms did not perform well. In their opinion, the sequential nature of
these hybrid algorithms: the second module is only able to make recommendations seeded by the
results of the first module. To address this problem, Ekstrand et al. \cite{ekstrand2010automatically}
propose to fuse the two steps by running a CF and a CBF
recommender in parallel and blending the resulting ranked lists. The first items on the combined
recommendation list are those items which appeared on both lists, ordered by the sum of
their ranks. Surprisingly, Collaborative Filtering outperforms
all hybrid algorithms in their experiments.

Gori et al. \cite{gori2006research} devised a random walk based method,
to recommend papers according to a small set of user selected relevant articles.
K\"u\c{c}\"uktun\c{c} et al. designed a personalized paper recommendation service, called theAdvisor\footnote{http://theadvisor.osu.edu/}~\cite{Kucuktunc13-ASONAM,Kucuktunc13-JCDL},
which allows a user to specify her search toward recent developments or traditional papers using a
direction-aware random walk with restart algorithm~\cite{Kucuktunc12-DBRank}.  The recommended papers
returned by theAdvisor are diversified by parameterized relaxed local maxima~\cite{Kucuktunc13-WWW}.
K\"u\c{c}\"uktun\c{c} et al. proposed sparse matrix ordering and partitioning techniques to accelerate citation such recommendation
algorithms~\cite{Kucuktunc12-ASONAM}.

Caragea et al.~\cite{caragea2013can} addressed the problem of citation recommendation
using singular value decomposition on the adjacency matrix associated with the citation graph
to construct a latent semantic space:
a lower-dimensional space where correlated papers can be  easily identified. Their experiments
on Citeseer show this approach achieves significant success compared with Collaborative
Filtering methods.
Wang et al.~\cite{Wang2011} proposes to include textual information to build an topic model of the papers and adds an additional latent variable to distinguish between the focus of a paper and the context of the paper.\looseness=-1

A typical related paper search scenario is that a user starts with a seed of one or more papers,
by reading the available text and searching related cited references. Sofia is a system
that automates this recursive process \cite{golshan2012sofia}.

The approach proposed by \cite{el2011beyond} returns a set of relevant articles by optimizing
a function based on a fine-grained notion of influence between documents; and also claim that, for paper recommendation, defining a query as a small set of known-to-be-relevant
papers is better than a string of keywords.

\emph{Manuscript based citation recommendation.}
%Citation recommendation by considering an unpublished manuscript as a query has also attracted
%much attention.
Stohman et al.~\cite{strohman2007recommending} examined the effectiveness of various
text-based and citation-based features on citation recommendation, they find that neither text-based nor
citation-based features performed very well in isolation, while text similarity alone achieves a
surprisingly poor performance at this task. He et al.~\cite{he2010context} considered the problem of
recommending citations for placeholder in query manuscripts and a proposed non-parametric probabilistic model to measure the relevance between a citation context and a candidate citation. To
reduce the burden on users, \cite{he2011citation}~proposed different models for automatically finding
citation contexts in an unlabeled query manuscript.

Recently, citation recommendation from heterogeneous network mining perspective has attracted more
attention. Besides papers, metadata such as authors or keywords are also considered as entities
in the graph schema. Two entities can be connected via different paths, called meta-paths, which usually
carry different semantic meanings. Many work build discriminative models for citation prediction and
recommendation based on meta-paths~\cite{yu2012citation,liu2014full,liu2014meta,ren2014cluscite}.

The vocabulary used in the citation context and in the content of papers are usually quite different.
To address this problem, some works propose to use translation model, which can bridge the gap
between two heterogeneous languages~\cite{lu2011recommending,huang2012recommending}. Based on
previous work~\cite{he2010context,he2011citation,huang2012recommending}, Huang et al. built a
citation recommendation system called RefSeer\footnote{http://refseer.ist.psu.edu/}~\cite{huang2014refseer} which perform both
topic-based global recommendations and citation-context-based local recommendations.

Based on the hypothesis that an author's published works constitute a clean signal of the
latent interests of a researcher, \cite{sugiyama2010scholarly}~examined the effect of modeling a
researcher's past works in recommending papers. Specifically, they first construct a user profile
based on her/his recent works, then rank candidate papers according to the content similarity
between the candidate and the user profile. Furthermore, in order to achieve a better representation
of candidate paper, \cite{sugiyama2013exploiting} exploit potential citation papers through the
use of collaborative filtering.

\section{Citation Recommendation}
\label{sec:problemstatement}

\subsection{Data Preparation}
To obtain a clean and comprehensive academic data set, we match
Microsoft Academic Graph\footnote{https://www.microsoft.com/en-us/research/project/microsoft-academic-graph/}~\cite{sinha2015overview},
CiteSeerX\footnote{http://citeseerx.ist.psu.edu/} and DBLP\footnote{http://dblp.uni-trier.de/xml/}~\cite{DBLP:journals/pvldb/Ley09}
datasets for their complementary advantages and derive a corpus of
Computer Science papers.
On one hand, Microsoft Academic Graph contains abundant information from various disciplines
but it is fairly noisy: some important attributes are missing or
wrong. In contrast, the records in DBLP are much more reliable
although it does not contain citation information. So we first merge MAG and DBLP records
through DOI and titles to get an academic citation graph (within the scope of Computer Science)
with rather clean metadata.\looseness=-1

On the other hand, CiteSeerX dataset indexes 2 million papers and provides full-texts in PDF
format which neither MAG or DBLP contains. We merge CiteSeerX and DBLP records through titles
and refine the result with other metadata, like published year.
This data set gives us for each paper the name of the authors, the
venue of publication, the title of the paper, full text (for about a
fifth of the papers), and citation information. We derived keywords
using KP-Miner~\cite{el2010kp} for those with full text and using non-stop words from titles for others. High
level statistics of this dataset is given in Table~\ref{tab:data-statistics}.

\begin{table}
\caption{Data Statistics}
\centering
\begin{tabular}{ccrrc}
 \hline
  \textbf{Attribute} & \textbf{Number} \\
  \rmfamily Papers & \rmfamily 2,035,246 \\
  \rmfamily Citations & \rmfamily 12,439,090 \\
  \rmfamily Papers with text & \rmfamily 374,999 \\
  \rmfamily Keywords & \rmfamily  195,989 \\
  \rmfamily P-K Edges & \rmfamily  14,779,751\\
\end{tabular}
\begin{tabular}{ccrrc}
 \hline
  \textbf{Attribute} & \textbf{Number} \\
  \rmfamily Authors & \rmfamily 1,208,641 \\
  \rmfamily P-A Edges & \rmfamily  5,977,884\\
  \rmfamily Venues & \rmfamily 9,777 \\
  \rmfamily P-V Edges & \rmfamily  2,035,246\\
  \hline
\end{tabular}
\label{tab:data-statistics}
\end{table}

\subsection{Problem Statement}

Let $G=(V,E)$ be the citation graph, with $n$ papers $V=\{v_1,\ldots,v_n\}$. In $G$, each
edge $e\in E$ represents a citation relationship between two papers.
We use $Ref(v)$ and $Cit(v)$ to denote the reference set of and citation set to $v$,
respectively. And $Adj(v)$ is used to denote the union of $Ref(v)$ and $Cit(v)$.

In this paper, we focus on citation recommendation problem assuming that researchers already
know a set of relevant papers.  Therefore, the task can be formalized as:

\emph{Citation Recommendation.} Given a set of seed papers $S$,
return a list of papers ranked by relevance to the ones in $S$.

\subsection{Algorithms}

\paragraph{CoCitation~\cite{boyack2010co}}
The number of cocitations is often used to measure the relevance between two papers.
In the citation recommendation scenario, cocitation ranks a candidate paper according
to the sum of the times it was cocited with papers in the seed set.
  \[
    R(x) = \begin{array}{lr}
        \sum_{s \in S}\sum_{v\text{ for }s,x \in Cit(v) }1
        \end{array}
  \]

\paragraph{CoCoupling~\cite{boyack2010co}}
CoCoupling is a complementary metric of cocitation. It counts the number of times that two
papers cite a same paper. Here, we use cocoupling to measure the relevance between the
candidate paper and seed papers according to the following formula:
  \[
    R(x) = \begin{array}{lr}
        \sum_{s \in S}\sum_{v\text{ for }s,x \in Ref(v) }1
        \end{array}
  \]

\paragraph{PaperRank~\cite{gori2006research} (PR)}
PaperRank is a biased random walk proposed to recommend papers based on citation
graph. In particular, the restarts from any paper will be distributed to only the seed papers. PR assumes
a random walker in paper $v$ continues to a neighbor with a damping factor $d$, and with probability $(1-d)$ it restarts
at one of the seed papers in $S$. The edges are followed proportionally to the weight of that edge $w_{ji}$ which is often set to 1, but can be set to the number of time $i$ is referenced by $j$.
$$R(v_{i})=(1-d)\frac{1}{S}+d\times{\sum_{v_{j}\in Adj(v_{i})}\frac{w_{ji}}{\sum_{v_{k}\in Adj(v_{j})}w_{jk}}R(v_{j})}$$

\paragraph{Collaborative Filtering~\cite{mcnee2002recommending} (CF)} has been proven to be an effective idea for most recommendation
problems. For citation recommendation, a ratings matrix is built using the adjacent matrix
of citation graph, where citing papers correspond to users and citations correspond to items.
A pseudo target paper that cites all seed papers is added to the matrix.
CF computes the $k$ neighborhoods that are top $k$ similar papers to the target paper.
Then each citation in neighbors is summed up with the count weighted by the similarity score.

\section{A First Evaluation}
\label{sec:globalperf}

In order to simulate the typical usecase where a researcher is writing
a paper and tries to find some more references, we design the random-hide
experiment. First of all, a query paper $q$ with 20 to 200
references and published between 2005 to 2010 is randomly (uniformly) selected from the dataset.  We then remove the
query paper $q$ and all papers published after $q$ from the citation
graph to simulate the time when the query paper was being
written. Instead of using hide-one
strategy~\cite{mcnee2002recommending,torres2004enhancing}, we randomly
hide 10\% of the references as hidden set. This set of hidden paper is used as ground
truth to recommend. The remaining (18 and 180 depending on $q$) papers are used as the set of seed papers.

Finally, to evaluate the
effectiveness of recommendation algorithm, we use \emph{recall@$k$},
the ratio of hidden papers appearing in top $k$ of the recommended
list. Table~\ref{tab:global_perf} shows the results of popular methods on
average recall for 2,500 independent randomly selected queries.
We call these scores global performance, as we will analyze the
common features of the hidden papers found by those methods to reveal
the bias behind the algorithms.

\begin{table}[tbp]
\caption{Global Performance}
\centering
\begin{tabular}{c|rrr}
 \hline
  \textbf{Method} & \textbf{Recall@10} & \textbf{Recall@20} & \textbf{Recall@50}\\\hline
  \rmfamily PaperRank & \rmfamily 0.234413  & \rmfamily 0.326096 & \rmfamily 0.471510\\
  \rmfamily CF & \rmfamily 0.191736  & \rmfamily 0.266961 & \rmfamily 0.391736\\
  \rmfamily CoCitation & \rmfamily 0.192626  & \rmfamily 0.267617 & \rmfamily 0.392197\\
  \rmfamily CoCoupling & \rmfamily 0.055778  & \rmfamily 0.088216 & \rmfamily 0.146737\\
  \hline
\end{tabular}
\label{tab:global_perf}
\end{table}

To analyze the performance of the algorithms, we investigate the local
structure of the citation graph. The citation projection graph of a
paper $p$ is the graph induced by the
papers cited by $p$~\cite{shi2010citing}. For a query paper, it is the
graph where the vertex set is composed of the seed papers and the
hidden papers, and the edge set is composed of the citations between
these vertices. The citation projection graph focuses on the cited
papers and the relationships among them; it is known to help
understanding the local pattern in the citation
graph~\cite{shi2010citing}.

\begin{figure}[tbp]
  \begin{center}
    \includegraphics[width=.7\linewidth]{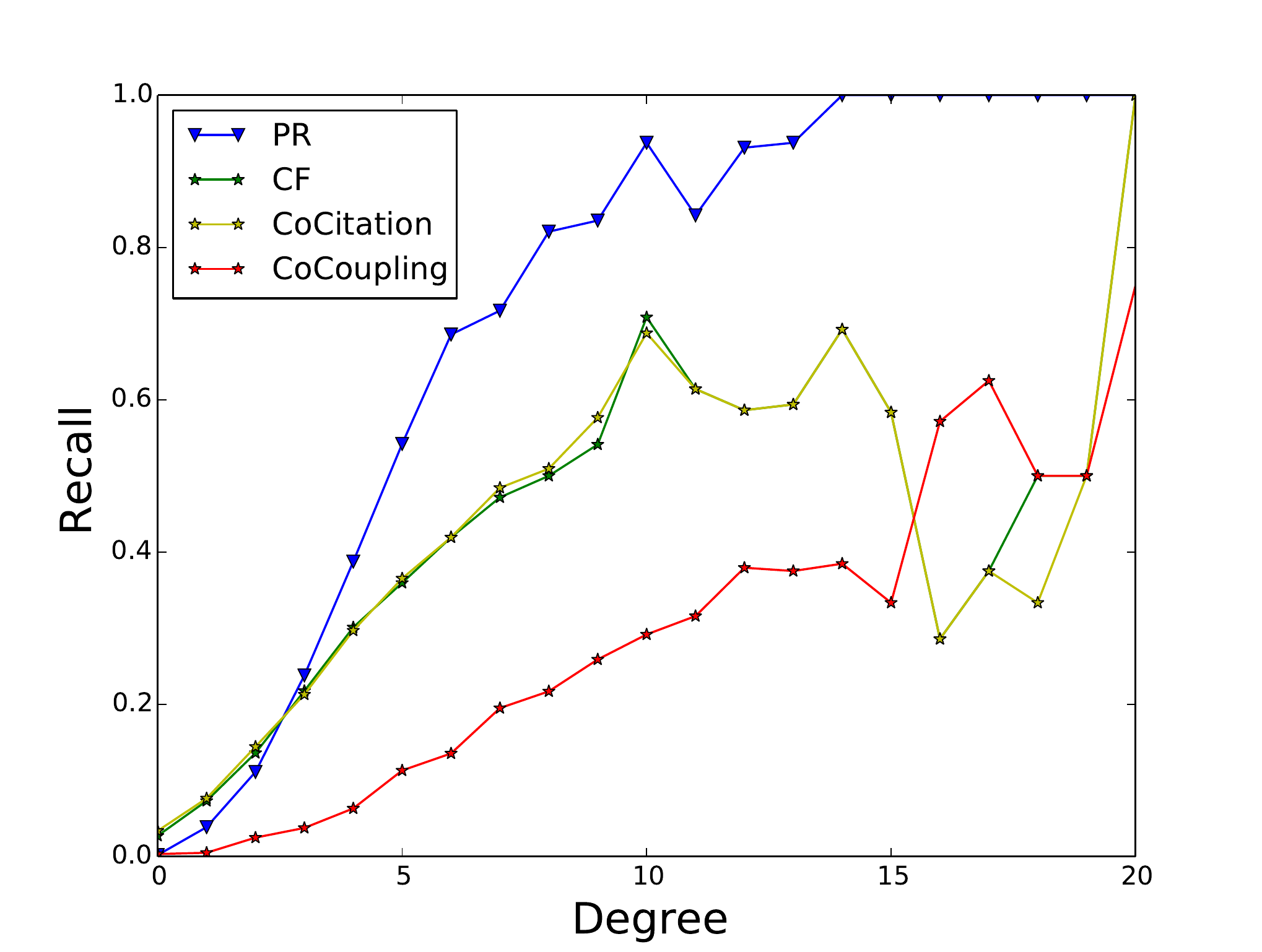}
  \end{center}
  \caption{Recall by proj-degree (top 10)}
  \label{fig:recall-by-projdegree}
\end{figure}

\begin{figure}[tbp]
  \begin{center}
    \includegraphics[width=.7\linewidth]{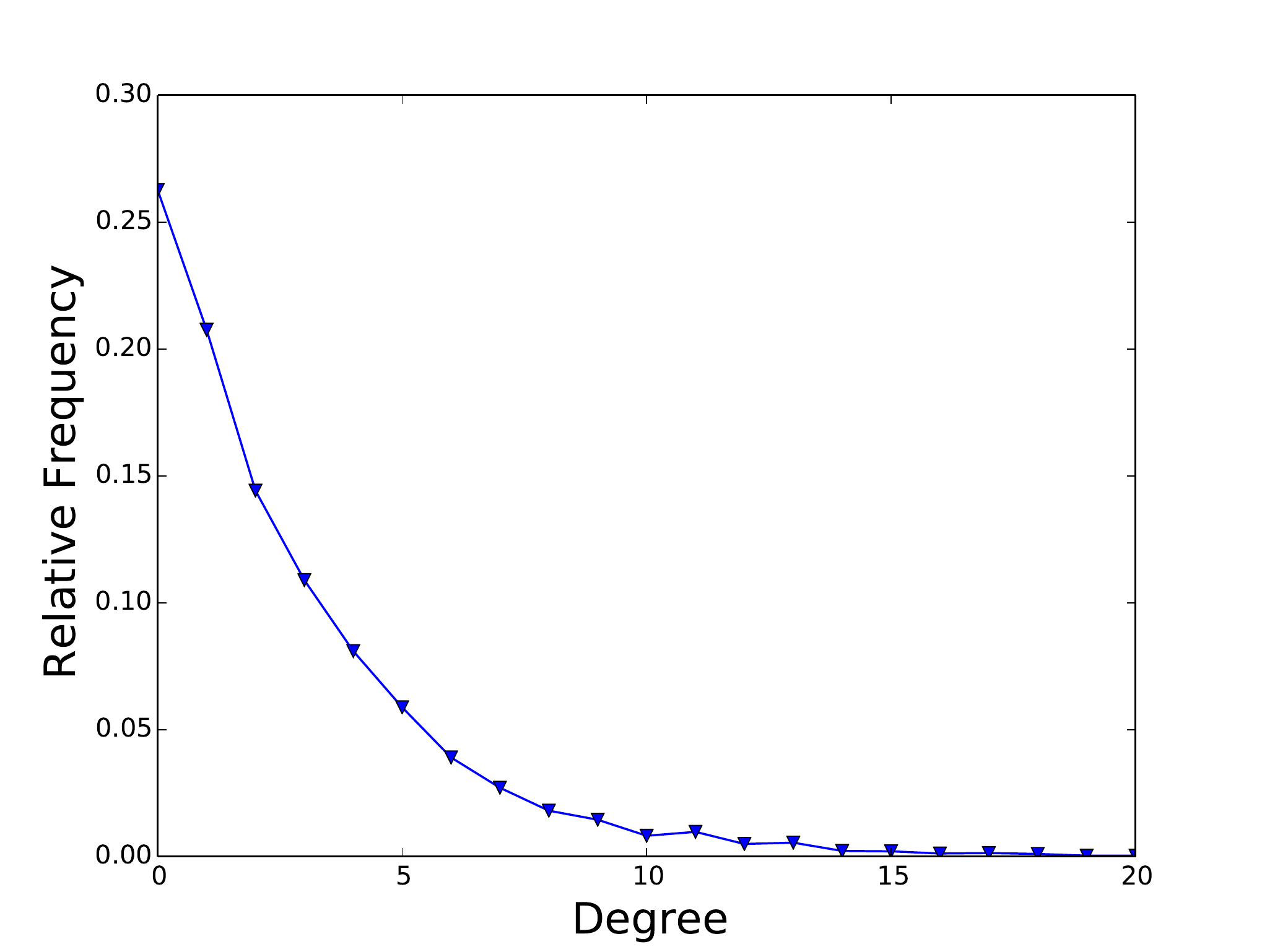}
  \end{center}
  \caption{Distribution of proj-degree (degree in the citation
    projection graph) of hidden papers.}
  \label{fig:projdegree-dist}
\end{figure}

We investigated the relations between various properties of the
projection graph and whether hidden papers were found or not.  We
identified that the degree of the hidden papers in the projection
graph, we call proj-degree, is a good indicator of whether the hidden
paper will be discovered or not. We computed the average recall@10
scores on hidden papers grouped by proj-degree and reported these
numbers in Figure~\ref{fig:recall-by-projdegree}. Popular graph based
methods are quite good at finding hidden papers that are highly
connected in the citation projection graph. But these methods
achieve poor performance on loosely connected ones. Unfortunately,
over 50\% of the hidden papers have a proj-degree of 2 or less. The distribution of proj-degree is shown in Figure~\ref{fig:projdegree-dist}.

\begin{figure}[tbp]
  \begin{center}
    \includegraphics[width=.7\linewidth]{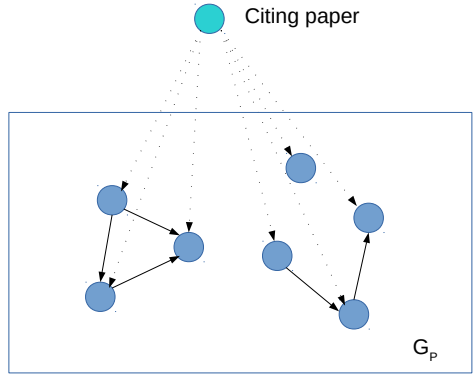}
  \end{center}
  \caption{Projection Graph}
  \label{fig:projection}
\end{figure}

\section{Finding Loosely Connected Papers}
\label{sec:finding-loosely}

\subsection{Are these papers random citation?}

The analysis of the last section shows  that popular methods are
good at finding papers that are highly connected in citation
projection graph. But they perform poorly on papers that are not well
connected in the citation projection graph despite they are the
majority. Therefore, we focus our analysis on loosely connected
papers.

The key question is why do authors cites these papers? According to
\cite{shi2010citing}, some papers create random reference across
various fields. This might sound reasonable to explain the fact that
these reference are loosely connected in the citation projection
graph. However, as Figure~\ref{fig:projdegree-dist} shows, about $50\%$ of cited papers
have one or no link to others. Therefore, we believe they must share
some common patterns or features with others cocited papers
that are not apparent in the citation graph.
We expect other features  such as authors,
venue, or keywords, convey helpful information.

A preliminary analysis of the metadata of loosely connected papers shows
that about $46\%$ of the papers connected to two of the seeds or less
share at least one common author with at least one of
the seed papers. $60\%$ of the loosely connected papers
appeared in the venue of one of the seed paper. And $95\%$ of the
loosely connected paper shared at least one keyword with one of the seed paper.
This indicates that the citations are not random citations; but
authors chose to cite them for reasons that are not clearly explained
by the citation graph.

% \begin{table}[!htbp]
% \centering
% \begin{tabular}{ccrrc}
%  \hline
%   \textbf{Attribute} & \textbf{Ratio} \\
%   \rmfamily Author & \rmfamily 0.4611 \\
%   \rmfamily Venue& \rmfamily 0.6080 \\
%   \rmfamily Keyword & \rmfamily 0.9536 \\
%   \hline
% \end{tabular}
% \label{tab:data}
% \caption{Ratio of loosely connected papers that have common attribute(s) with at least one of seed papers.}
% \end{table}

\subsection{Algorithms using Metadata}

Based on above analysis, we explore different approaches that use metadata for citation recommendation.
we firstly define a group of paper ranking schemes based on meta path in the bibliographic network.
Then we propose two random walk based methods to examine the ability of metadata for identifying loosely connected papers.
One is based on the metadata themselves, and we combine the metadata and citation graph in the other one.

\subsubsection{MetaPath}

Recently, similarity search among the same type of entities in
heterogeneous networks has attracted more attention. Intuitively,
two entities are similar if they are linked by many paths
in the network. However, most existing similarity measures are
defined for homogeneous networks. Therefore, meta path-based
similarity is proposed~\cite{sun2011pathsim}.

A meta path is a path defined on the heterogeneous network schema, and
is denoted in the form of $O_{1} \rightarrow^{R_{1}}O_{2} \rightarrow^{R_{2}}... \rightarrow^{R_{l}}O_{l+1} $, which defines a composite relation $R$ between type $O_{1}$ and $O_{l+1}$.

In a heterogeneous network, two entities can be connected via
different meta paths. For example, in bibliographic network, two authors can be connected
via "author-paper-author"path, "author-paper-venue-paper-author" path, and so on.
Different meta paths usually carry different semantic meanings.

For citation recommendation problem, we are looking for papers that
are relevant to the seed papers. To measure the relevance between a pair of
papers, 5 basic meta paths are defined:
\begin{itemize}
\item $Paper - Author - Paper(PAP)$: Two papers may be relevant if they share a author
\item $Paper - Venue - Paper(PVP)$: Two paper may be relevant if they are published at the same venue.
\item $Paper - Keyword - Paper(PKP)$: Two paper may be relevant if they share a keyword.
\item $Paper\rightarrow Paper\leftarrow Paper(PCiP)$: Two paper may be relevant if they share a citation.
\item $Paper\leftarrow Paper\rightarrow Paper(PCoP)$: Two paper may be relevant if they are cited by the same paper.
\end{itemize}

Given a paper-to-paper meta path, several similarity measures can be defined
according to the path instances between them following the meta path. A straightforward
measure will be:

\textbf{PathCount Measure:} Given a meta path $\mathcal{P}$ and a pair of papers $x$ and $y$, the similarity between
them is defined as:
$$ PathCount(x,y,\mathcal{P}) = |\{p_{x\rightarrow y}:p_{x\rightarrow y}\in \mathcal{P}\}|$$

Essentially, PathCount is the number of path instances $p$ between $x$ and $y$.
This kind of similarity always favor entities with large degrees.
Therefore, Sun et al.\cite{sun2011pathsim} propose a new meta path
based similarity measure, called PathSim, which tries to capture
the subtlety of peer similarity:

\textbf{PathSim Measure:} Given a symmetric meta path $\mathcal{P}$, the similarity between two entities of the same type $x$ and $y$ is:
$$PathSim(x,y,\mathcal{P})=\frac{2 \times |\{p_{x\rightarrow y} : p_{x\rightarrow y}\in \mathcal{P}\}| }{|\{p_{x\rightarrow x} : p_{x\rightarrow x}\in \mathcal{P}\}|+|\{p_{y\rightarrow y} : p_{y\rightarrow y}\in \mathcal{P}\}|}$$

where $p_{x\rightarrow y}$ is a path instance between $x$ and $y$.
$p_{x\rightarrow x}$ is that between $x$ and $x$, and $p_{y\rightarrow y}$ is that between $y$ and $y$.

The intuition behind PathSim is that two similar peer entities should
not only be strongly connected, but also share comparable visibility,
where the visibility is defined as the number of path instances.
between themselves.

\begin{figure}[tbp]
  \begin{center}
    \includegraphics[width=.7\linewidth]{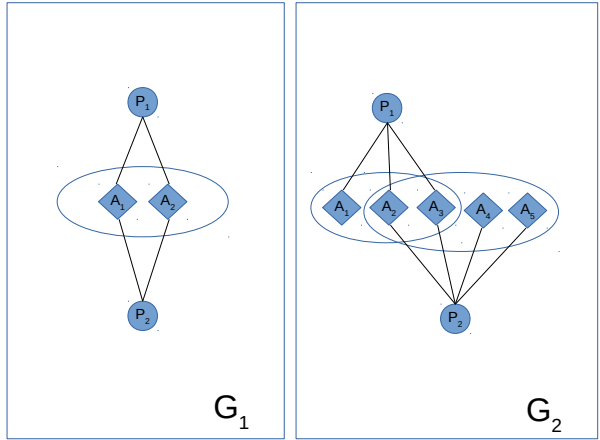}
  \end{center}
  \caption{Meta path examples: Paper-Author-Paper}
  \label{fig:PAP}
  \vspace{-2em}
\end{figure}

Figure~\ref{fig:PAP} shows two examples induced from the bibliographic
network. In $G_{1}$, both $P_{1}$ and $P_{2}$ are written by the
same two authors, while in $G_{2}$, $P_{1}$ also shares two common authors
with $P_{2}$ but $P_{1}$ and $P_{2}$ have 3 and 4 authors in total respectively.
The $PathCount(P_{1},P_{2},PAP)$ scores between $P_{1}$ and $P_{2}$ in $G_{1}$
and $G_{2}$ are the same since there are 2 PAP paths for both examples.
However, the $PathSim(P_{1},P_{2},PAP)$ scores are different: for $G_{1}$,
$PathSim(P_{1},P_{2},PAP)=\frac{2\times 2}{2+2}=1$ and for $G_{2}$, $PathSim(P_{1},P_{2},PAP)=\frac{2\times 2}{3+4}=0.57$.

Based on above similarity measures, the relevance between a candidate paper $x$ and seed papers
can be denoted as:
$$Score(x,Seed,\mathcal{P}) = \sum_{s\in Seeds}\frac{Score(x,y,\mathcal{P})}{|\{s:s\in Seeds\}|}$$
where $Score$ function is either $PathCount$ or $PathSim$.

Now we have 5 different paper-to-paper meta paths and 2 meta path-based measures.
Theoretically, there will be $5\times2$ ways to rank the candidate papers.
In particular, $PathCount_{PCiP}$ is essentially the CoCoupling method and
the same for $PathCount_{PCoP}$ and the CoCitation method. Besides, as a paper-to-venue
is always a one-to-one pair, $PathCount_{PVP}$ and $PathCSim_{PVP}$ will be the same thing.
Therefore, we have 7 meta path based ranking methods, namely:
$PathCount_{PAP}$, $PathCount_{PKP}$ and $PathSim_{PAP}$, $PathSim_{PVP}$, $PathSim_{PKP}$, $PathSim_{PCiP}$ and $PathSim_{PCoP}$.

%\begin{table}[tbp]
%\caption{MetaPath}
%\centering
%\begin{tabular}{|c|p{5cm}|}
% \hline
%  \textbf{MetaPath} & \textbf{Description} \\\hline
%  \rmfamily $P-A-P$ & \rmfamily A paper may be relevant if it shares a author with a seed paper\\\hline
%  \rmfamily $P-V-P$ & \rmfamily A paper may be relevant if it is published at the same venue as the seed paper  \\\hline
%  \rmfamily $P-K-P$ & \rmfamily A paper may be relevant if it shares a keyword with a seed paper\\\hline
%  \rmfamily $P\rightarrow P\leftarrow P$ & \rmfamily   \\\hline
%  \rmfamily $P\leftarrow P\rightarrow P$ & \rmfamily   \\
%  \hline
%\end{tabular}
%\label{tab:glob-2}
%\end{table}
\subsubsection{LOGAVK}
In order to compute the similarity between one paper to a set of other papers, we build attribute graphs for author, venue and keyword respectively.
Let us take author as example, we first define an undirected weighted graph of authors where an edge represents the number of papers two authors
have written together. Then we normalize the adjacent matrix of this graph as $M^{AA}$, where $A$ is the set of authors. Once the
graph is constructed, we can measure the similarity between a candidate author and the authors of seed papers by random walk as follows:

 \[
  R^{A} =
  \begin{cases}
    \alpha M^{AA}  R^{A}  + (1-\alpha)\frac{1}{S} & \text{For authors of seed papers} \\
    \alpha M^{AA}  R^{A}  & \text{otherwise}
  \end{cases}
 \]
The keyword graph $M^{KK}$ and venue graph $M^{VV}$  are constructed and the similarity score $R^{V}$ and $R^{K}$ are computed in the same way.
LogAVK recommends the loosely connected papers according to the summation of the similarity scores of authors, venue and keywords with corresponding
seed papers.
$$ Score_{LogAVK} =logR^{A}+logR^{V}+logR^{K}$$

\subsubsection{Combining Citation graph and Metadata}

Aiming to combine the citation information and metadata information, we build bipartite graphs with two kinds of nodes: papers and metadata.
A random walk algorithm passes information back and forth between the papers and the metadata.
Taking the paper-author graph as an example, the vector of paper scores is denoted by $R^{P}$ and the vector of author scores is denoted
by $R^{A}$. The scores of authors is computed by:

$$R^{A} = M^{AP} R^{P}$$

which means an author score is collected from the papers she published.
Some of the paper scores are transferred between papers within the citation graph:
$$R^{P}_{1} = M^{PP} R^{P} $$
And a paper also collects scores from its authors:
$$R^{P}_{2} = M^{PA} R^{A} $$
A paper in the seed set $S$ also receives scores by random jumping from others.
$$R^{P}_{3} = \frac{1}{S} $$
The final score of a paper is the weighted sum of above parts.
\[
R^{P}_{A} =
\begin{cases}
  \alpha R^{P}_{1} +\beta R^{P}_{2} +(1-\alpha-\beta)R^{P}_{3} & \text{for seed papers} \\
  \alpha R^{P}_{1}+\beta R^{P}_{2}   & \text{otherwise}\\
\end{cases}
\]
where $\alpha$ ($\beta$. resp.) is the fraction of the rank following a citation edge (an author edge, resp.). In the experiments, we set $\alpha = .65$, $\beta = .2$.

We will refer to any method that combines the citation information and
a metadata in this manner as C+X. In particular, C+A will denote
combining citation and authorship; C+K will denote
combining citation and keyword; and C+V will denote
combining citation and venue.

\subsection{Evaluation}

\paragraph{General performance}
We evaluate the general effectiveness of the recommendation
algorithms using \emph{recall@$k$}, the ratio of hidden papers
appearing in top $k$ of the recommended list. Table~\ref{tab:glob-2} shows the
results of popular methods and the methods on average recall for
2,500 independent queries.

\begin{table}[tbp]
\caption{Global Performance}
\centering
\begin{tabular}{c|rrr}
 \hline
  \textbf{Method} & \textbf{Recall@10} & \textbf{Recall@20} & \textbf{Recall@50}\\\hline
  \rmfamily PaperRank & \rmfamily 0.234413  & \rmfamily 0.326096 & \rmfamily 0.471510\\
  \rmfamily CF & \rmfamily 0.191736  & \rmfamily 0.266961 & \rmfamily 0.391736\\
  \rmfamily C+A & \rmfamily 0.230617  & \rmfamily 0.318206 & \rmfamily 0.463204\\
  \rmfamily C+V & \rmfamily 0.230531  & \rmfamily 0.323125 & \rmfamily 0.461898\\
  \rmfamily C+K & \rmfamily 0.231308  & \rmfamily 0.315485 & \rmfamily 0.461507\\
  \rmfamily LogAVK & \rmfamily 0.053934  & \rmfamily 0.084001 & \rmfamily 0.129175\\
  \rmfamily PathCount\_PAP & \rmfamily 0.053291  & \rmfamily 0.079318 & \rmfamily 0.125437\\
  \rmfamily PathCount\_PKP & \rmfamily 0.031268  & \rmfamily 0.050866 & \rmfamily 0.083506\\
  \rmfamily PathSim\_PAP & \rmfamily 0.053374  & \rmfamily 0.079897 & \rmfamily 0.125602\\
  \rmfamily PathSim\_PVP & \rmfamily 0.003057  & \rmfamily 0.005231 & \rmfamily 0.010377\\
  \rmfamily PathSim\_PKP & \rmfamily 0.031662  & \rmfamily 0.051366 & \rmfamily 0.098917\\
  \rmfamily PathSim\_PCiP & \rmfamily 0.061189  & \rmfamily 0.095165 & \rmfamily 0.158142\\
  \rmfamily PathSim\_PCoP & \rmfamily 0.192168  & \rmfamily 0.269849 & \rmfamily 0.396291\\
  \hline
\end{tabular}
\label{tab:glob-2}
\end{table}

The results show that the C+X methods do not perform
quite as well as PaperRank; while the performance of logAVK
is lower than that of PaperRank by a factor of about 4.

It seems methods that merely rely on metadata are not working well for
citation recommendation task. Nevertheless, we can still conclude some
interesting findings from those methods: Author path and keyword path
are more useful than venue path; Pathsim tends to be a better meta path
measure comparing with PathCount. In the following sections, we keep
the best performed meta path, PathSim\_PCoP, for further study.

\emph{Performance by proj-degree.}
In order to evaluate the ability to recommend papers with a particular
degree in the citation projection graphs, we design the second
experiment. We define \emph{recall@$k$} for {$\delta=\Delta$} as the ratio
of hidden papers with proj-degree $d$ to seeds papers appearing in top
$k$ of the recommended list, where only the papers with proj-degree
$\Delta$ to seeds papers are considered as candidates\footnote{We call it
  property proj-degree for simplicity. Indeed the method would need to
  know which are the hidden paper to do the filtering on
  proj-degree. We mean degree to the seed, which differs from the real
  proj-degree by the number of connections to the unknown hidden.}. The
results are shown in Table~\ref{tab:recall-delta}.

\begin{table}[tbp]
\caption{Performance by proj-degree: Recall@10}
%\centering
\resizebox{\linewidth}{!}{
\begin{tabular}{l|rrrrrr}
 \hline
  \textbf{Method} & \textbf{$\delta=0$} & \textbf{$\delta=1$} & \textbf{$\delta=2$} & \textbf{$\delta=3$} & \textbf{$\delta=4$} & \textbf{$\delta=5$}\\\hline
  \rmfamily Cocitation & \rmfamily 0.10465  & \rmfamily 0.20760  & \rmfamily 0.46879  & \rmfamily 0.73310  & \rmfamily 0.88773 & \rmfamily 0.94630\\
  \rmfamily Cocoupling & \rmfamily 0.01723  & \rmfamily 0.03577  & \rmfamily 0.11962  & \rmfamily 0.25298  & \rmfamily 0.46489 & \rmfamily 0.64429\\
  \rmfamily Co. Filtering& \rmfamily 0.09914  & \rmfamily 0.20051  & \rmfamily 0.47150  & \rmfamily 0.73821  & \rmfamily 0.89120 & \rmfamily 0.94630\\
  \rmfamily PaperRank & \rmfamily 0.12172  & \rmfamily 0.20284  & \rmfamily 0.50463  & \rmfamily \textbf{0.76831}  & \rmfamily 0.91358 & \rmfamily 0.96979\\
  \rmfamily C+A & \rmfamily 0.11193  & \rmfamily 0.20719  & \rmfamily \textbf{0.51515}  & \rmfamily 0.76490  & \rmfamily 0.91319 & \rmfamily \textbf{0.97147}\\
  \rmfamily C+V & \rmfamily 0.11544  & \rmfamily 0.18023  & \rmfamily 0.49932  & \rmfamily 0.76661  & \rmfamily \textbf{0.92168} & \rmfamily \textbf{0.97147}\\
  \rmfamily C+K & \rmfamily \textbf{0.14160}  & \rmfamily 0.17829  & \rmfamily 0.50260  & \rmfamily 0.76008  & \rmfamily 0.91242 &\rmfamily \textbf{0.97147}\\
  \rmfamily logAVK & \rmfamily 0.02394  & \rmfamily 0.10287  & \rmfamily 0.30427  & \rmfamily 0.55820  & \rmfamily 0.80671 & \rmfamily 0.91778\\
  \rmfamily PathSim\_PCoP & \rmfamily 0.10488  & \rmfamily \textbf{0.20875}  & \rmfamily 0.47083  & \rmfamily 0.73461  & \rmfamily 0.88218 & \rmfamily 0.94630\\
  \hline
\end{tabular}
}\label{tab:recall-delta}
\end{table}

For particular values of proj-degree, the combined methods (C+X)
outperform current methods. One can easily see that most methods
perform well on high proj-degrees. Indeed, there are few vertices that are
very connected with the seed papers. So any reasonable algorithm will
find most of them. It is on lower proj-degrees (0, 1, and 2) that
the algorithms start finding less than 50\% of the hidden papers.

\begin{figure*}[tbp]
\centering
\begin{subfigure}{0.51\linewidth}
  \centering
  \includegraphics[width=\textwidth]{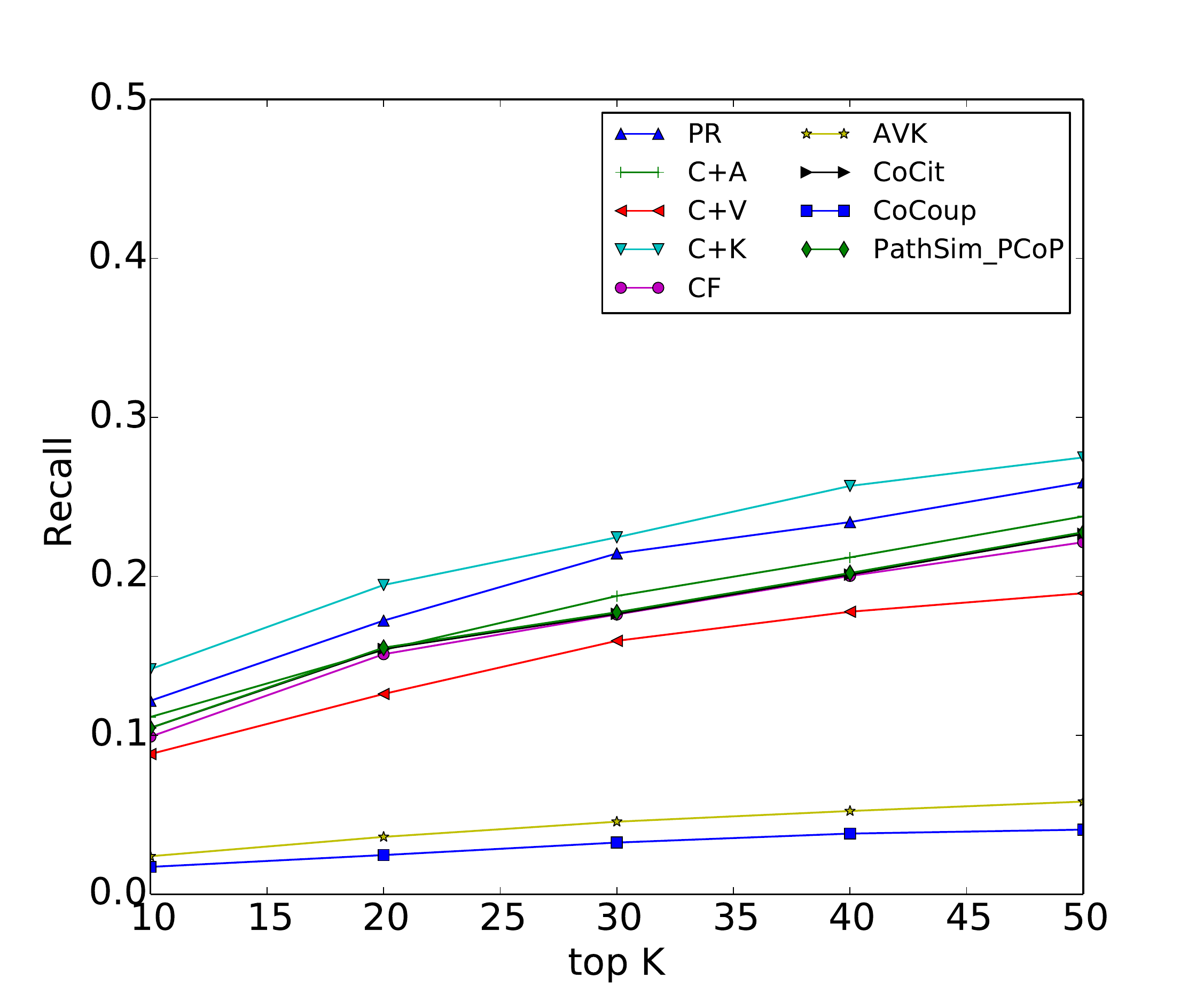}
  \caption{$\delta=0$}
\end{subfigure}
\begin{subfigure}{0.49\linewidth}
  \centering
  \includegraphics[width=\textwidth]{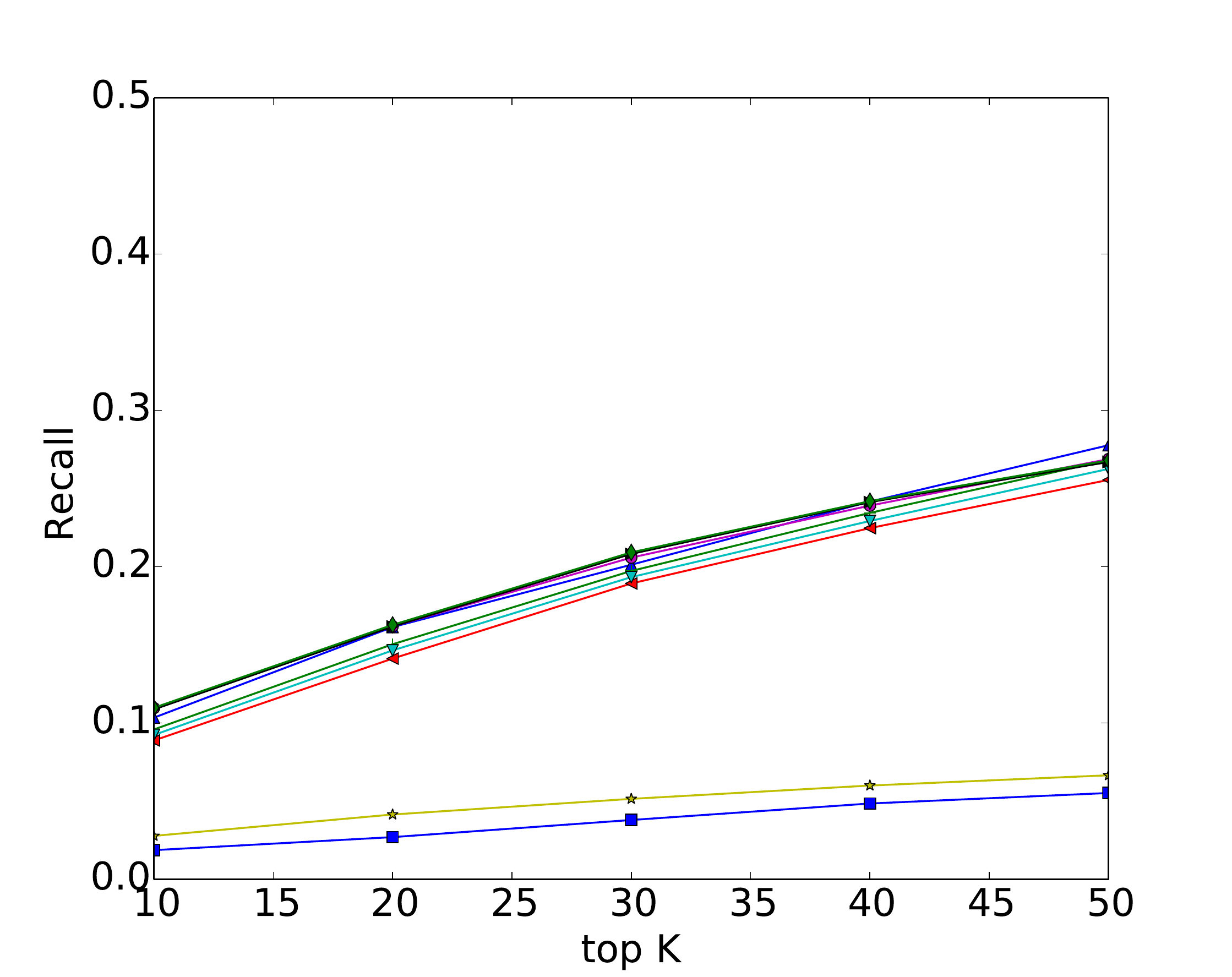}
  \caption{$\delta \leq 1$}
\end{subfigure}
\begin{subfigure}{0.49\linewidth}
  \centering
  \includegraphics[width=\textwidth]{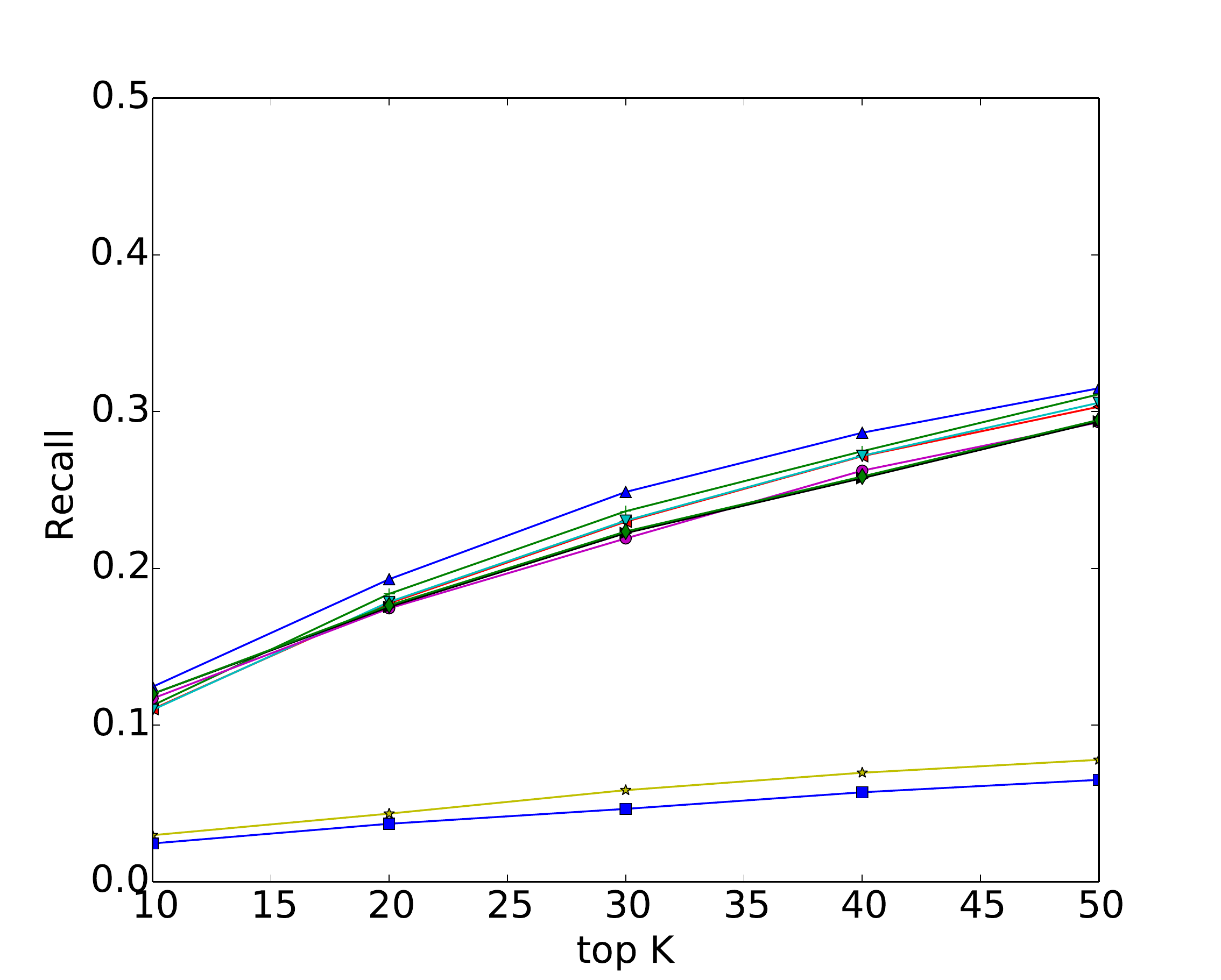}
  \caption{$\delta \leq 2$}
\end{subfigure}

% \begin{subfigure}{0.3\textwidth}
%   \centering
%   \includegraphics[width=\textwidth]{plots/deg3.pdf}
%   \caption{$\delta<=3$}
% \end{subfigure}

% \begin{subfigure}{0.3\textwidth}
%   \centering
%   \includegraphics[width=\textwidth]{plots/deg4.pdf}
%   \caption{$\delta<=4$}
% \end{subfigure}

% \begin{subfigure}{0.3\textwidth}
%   \centering
%   \includegraphics[width=\textwidth]{plots/deg5.pdf}
%   \caption{$\delta<=5$}
% \end{subfigure}
\caption{Performance Comparison for low degree}
\label{fig:recall-low-degree}
\vspace{-1em}
\end{figure*}

Figure~\ref{fig:recall-low-degree} shows the evolution of the recall
when the number of returned papers varies for three definitions
of low proj-degree ($\delta = 0$, $\delta \leq 1$, $\delta \leq 2$). The
performance of the algorithms for $\delta\leq 1$ and $\delta \leq 2$
are similar: all graph based methods perform about the same (except
cocoupling). logAVK performs significantly worse.
For completely disconnect papers ($\delta = 0$), the graph based
algorithms exhibit more difference. And in particular, C+K performs better
than all other tested algorithms, besting PaperRank by $.02$. This
indicates that metadata help finding loosely connected paper.

\section{How the quality of metadata affects the performance}
\label{sec:keywords}

As we can see in above section, metadata can help us to find some
relevant results from a different perspective on the data. However,
it is difficult to guarantee the accuracy of metadata since they are
automatically extracted from multi-source data in most cases.

In our experiments, authors and venues are rather clean because they
are extracted from DBLP, which is a well-maintained high-quality database,
and matched with Microsoft Academic Graph Metadata. Therefore, we focus on
keywords in this section. In specific, we discuss different keywords extraction
strategies and conduct a empirical study to show their affects on the performance.

Keywords are the words that provide a brief and precise description for
a document. Even though there are so much work on automatic keyword
extraction problem, state-of-the-art methods can not get satisfying
performance, which means this problem is till far from being solved.
Most popular methods rely on part-of-speech tags, which is very expensive
for large corpus. Here we use KP-Miner~\cite{el2010kp}, which is fast and proven to be
effective for keywords extraction task from scientific documents~\cite{Kim:10b}.

However, all existing methods are designed for extracting keywords from
text-abundant documents. In our case, only 374,999 of 2,035,246 papers contain
full text information, while the rest of them just have titles. And our target is to
derive keywords for each paper. So we propose the following three strategies and
show their performance on citation recommendation task.

\begin{itemize}
\item \emph{KP-Miner+Titles}  This is the default strategy we used in above experiments.
In specific, we derive keywords using KP-Miner for those with full text and using non-stop words from titles for others.
\item \emph{KP-Miner+Propagation} We derive keywords using KP-Miner for those with full text and propagate the derived
keywords on citation graph through label propagation algorithm.
\item \emph{Titles} We derive keywords using non-stop words from titles for all papers.
\end{itemize}

\begin{table}[tbp]
\caption{Different keyword extraction strategies}
\centering
\begin{tabular}{c|rrr}
 \hline
  \textbf{Method} & \textbf{Recall@10} & \textbf{Recall@20} & \textbf{Recall@50}\\\hline
  \rmfamily KP-Miner+Titles & \rmfamily 0.231308 & \rmfamily 0.315485 & \rmfamily 0.461507\\
  \rmfamily KP-Miner+Propagation & \rmfamily 0.214116 & \rmfamily 0.295392  & \rmfamily 0.429822\\
  \rmfamily Titles & \rmfamily 0.231203 & \rmfamily 0.314752  & \rmfamily 0.460039\\
  \hline
\end{tabular}
\label{tab:C+K}
\end{table}

Table~\ref{tab:C+K} shows the global performance of C+K with different strategies on the citation recommendation task.
In particular, \emph{KP-Miner+Propagation} is slightly worse than \emph{KP-Miner+Titles}, the main reason is that
propagating keyword information on citation graph will mislabel some papers. Surprisingly, the strategy that
derives keywords from titles for all papers is as good as \emph{KP-Miner+Titles}. To keep consistency, we
still use \emph{KP-Miner+Titles} strategy for C+K in the rest of this paper.

\section{On the usefulness of different algorithms}
\label{sec:usefulness}

\subsection{Difference between methods}

Looking at recall numbers gives a single perspective on the usefulness
of the methods. Recall numbers tell us how the algorithms perform on
some particular test. While informative to pick a single ``best''
algorithm, a user wants to explore a dataset and see it through
different lenses.

Table~\ref{tab:diff-delta2} allows us to understand how similar the
sets recommended by the algorithms are for loosely connected
papers. The diagonal shows the number of hidden papers that were found
in the top-10 by a particular algorithm, while an off diagonal entries
shows the number of paper found by the algorithm of the row and that
were not found by the algorithm on the column. For instance,
Cocitation recommended correctly 408 papers but only 393 of those were
not correctly identified by CoCoupling.

\begin{table}[tbp]
\caption{Differences between the top-10 sets ($\delta \leq 2$)}
\centering
\resizebox{\linewidth}{!}{
  \begin{tabular}{ccccccccccccc}
    \hline
    \textbf{}                 & \textbf{CoCit}   & \textbf{CoCoup}  & \textbf{CF}    & \textbf{PR}   & \textbf{C+A}  & \textbf{C+V}  & \textbf{C+K}  & \textbf{logAVK} & \textbf{PathSim\_PCoP} \\
    \rmfamily \textbf{CoCit}    & \rmfamily \textbf{408} & \rmfamily 393 & \rmfamily 45 & \rmfamily 245 & \rmfamily 284 & \rmfamily 289 & \rmfamily 293 & \rmfamily 389 & \rmfamily 10\\
    \rmfamily \textbf{CoCoup}    & \rmfamily 62 & \rmfamily \textbf{77} & \rmfamily 60  & \rmfamily 57 & \rmfamily 64 & \rmfamily 66 & \rmfamily 65 & \rmfamily 75 & \rmfamily 62\\
    \rmfamily \textbf{CF}     & \rmfamily 33 & \rmfamily 379 & \rmfamily \textbf{396} & \rmfamily 229 & \rmfamily 268 & \rmfamily 274 & \rmfamily 278 & \rmfamily 378 & \rmfamily 35\\
    \rmfamily \textbf{PR}     & \rmfamily 253 & \rmfamily 396 & \rmfamily 249  & \rmfamily \textbf{416} & \rmfamily 83 & \rmfamily 79 & \rmfamily 84 & \rmfamily 396 & \rmfamily 254\\
    \rmfamily \textbf{C+A}    & \rmfamily 259 & \rmfamily 370 & \rmfamily 255  & \rmfamily 50  & \rmfamily \textbf{383} & \rmfamily 58 & \rmfamily 62  & \rmfamily 362 & \rmfamily 260\\
    \rmfamily \textbf{C+V}    & \rmfamily 253 & \rmfamily 361 & \rmfamily 250   & \rmfamily 35  & \rmfamily 47  & \rmfamily \textbf{372} & \rmfamily 17  & \rmfamily 356 & \rmfamily 253\\
    \rmfamily \textbf{C+K}    & \rmfamily 256 & \rmfamily 359 & \rmfamily 253  & \rmfamily 39  & \rmfamily 50  & \rmfamily 16 & \rmfamily \textbf{371} & \rmfamily 356 & \rmfamily 256\\
    \rmfamily \textbf{logAVK} & \rmfamily 94 & \rmfamily 111 & \rmfamily 95   & \rmfamily 93  & \rmfamily 92  & \rmfamily 97 & \rmfamily 98  & \rmfamily \textbf{113} & \rmfamily 94\\
    \rmfamily \textbf{PathSim\_PCoP} & \rmfamily 13 & \rmfamily 396 & \rmfamily 50   & \rmfamily 249  & \rmfamily 288  & \rmfamily 292 & \rmfamily 296  & \rmfamily 392 & \rmfamily \textbf{411} \\
    \hline
  \end{tabular}
}
\label{tab:diff-delta2}
\end{table}

This table allows us to understand that PaperRank, C+A, C+K, and C+V
essentially identify the same papers. Indeed each set is composed of
about 400 papers, but the difference between these sets is smaller
than 100 papers and often smaller than 50 papers. Similarly, Cocitation
and Collaborative Filtering both find about 400 papers, but only about 40 of these papers
are actually different.

\begin{figure}[tbp]
  \begin{subfigure}{0.48\linewidth}
    \centering
    \includegraphics[width=\textwidth]{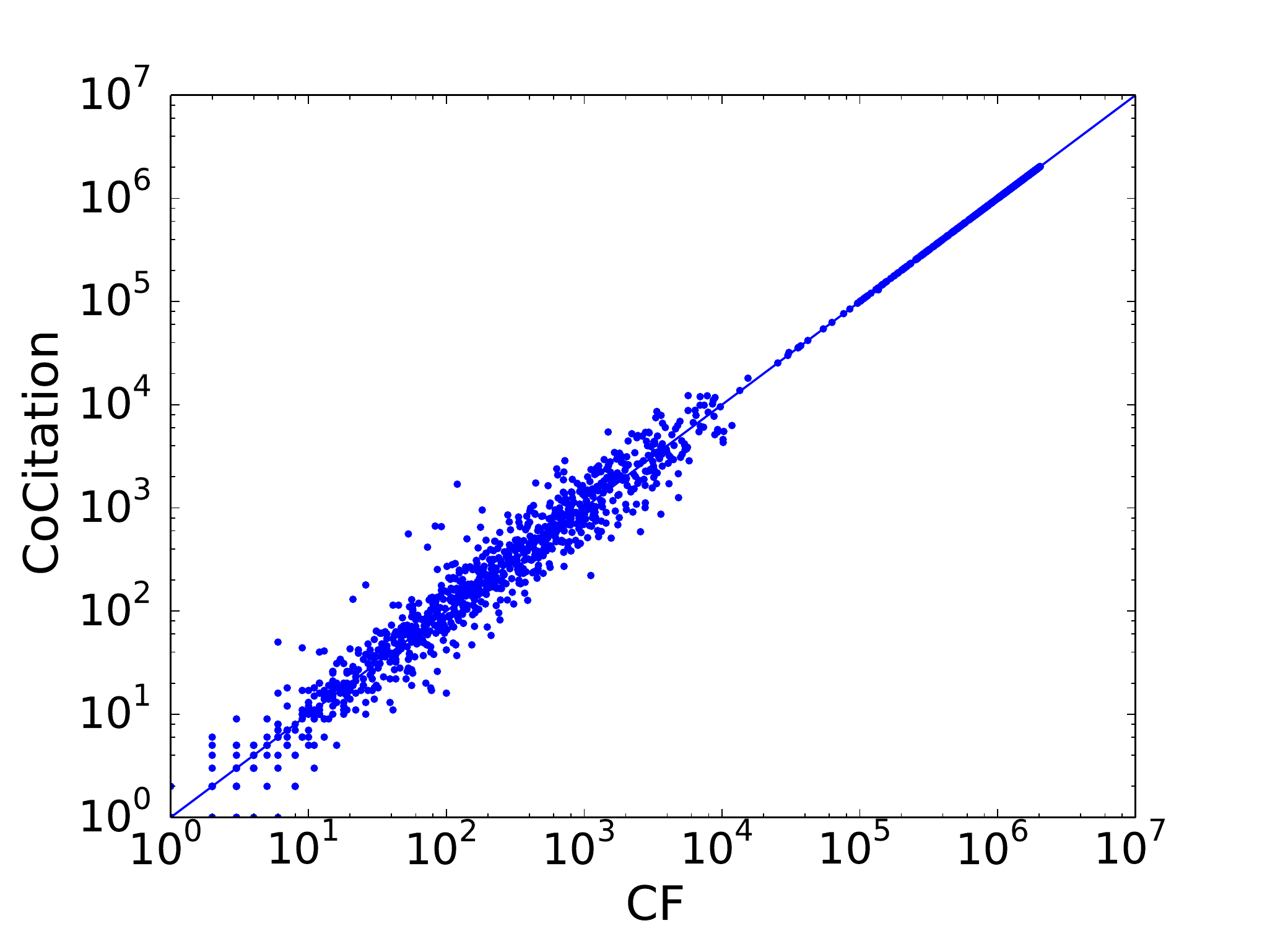}
    \caption{CF vs Cocitation}
  \end{subfigure}
\hfill
  \begin{subfigure}{0.48\linewidth}
    \centering
    \includegraphics[width=\textwidth]{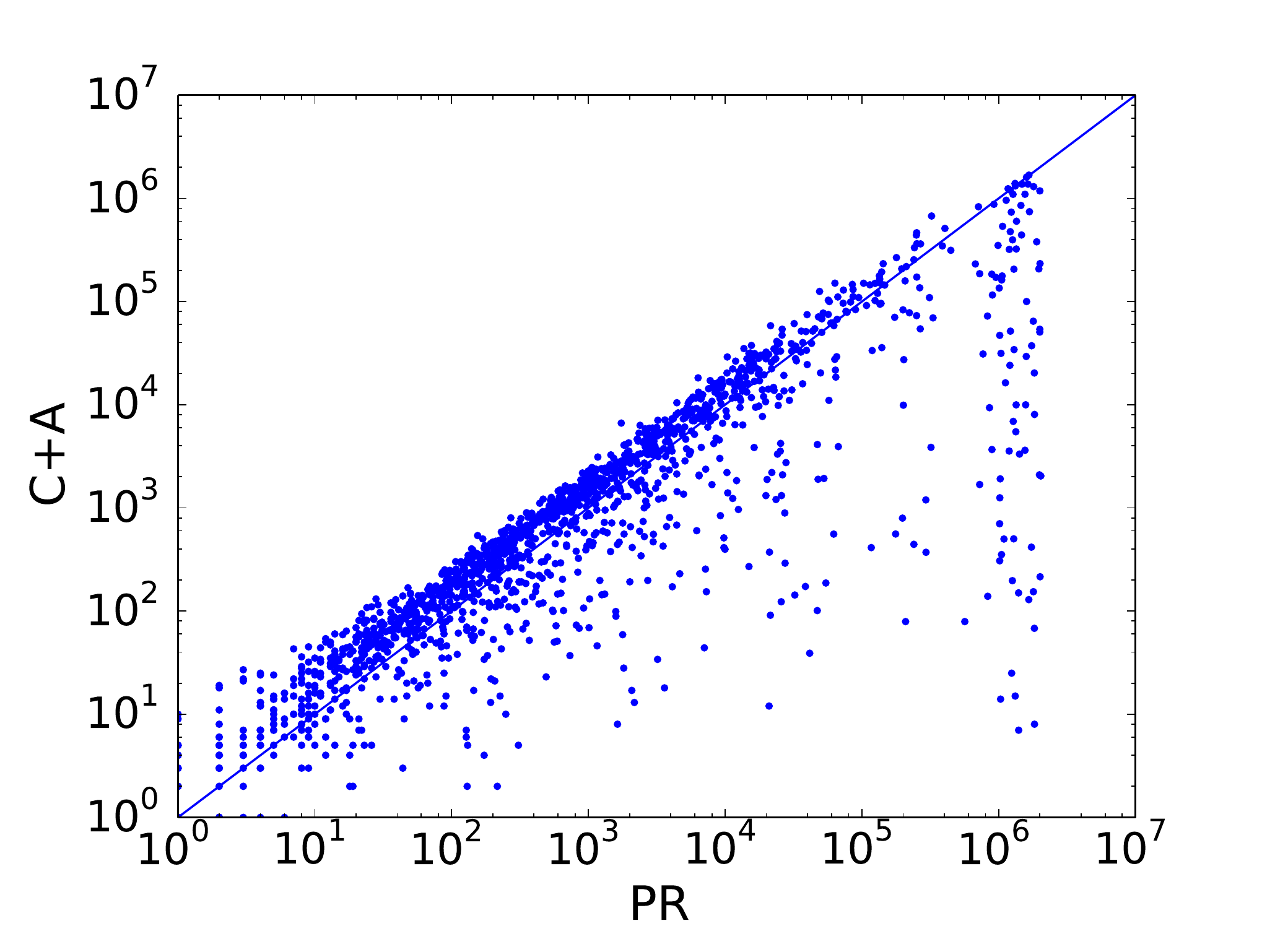}
    \caption{PR vs C+A}
  \end{subfigure}
  \hfill
  \begin{subfigure}{0.48\linewidth}
    \centering
    \includegraphics[width=\textwidth]{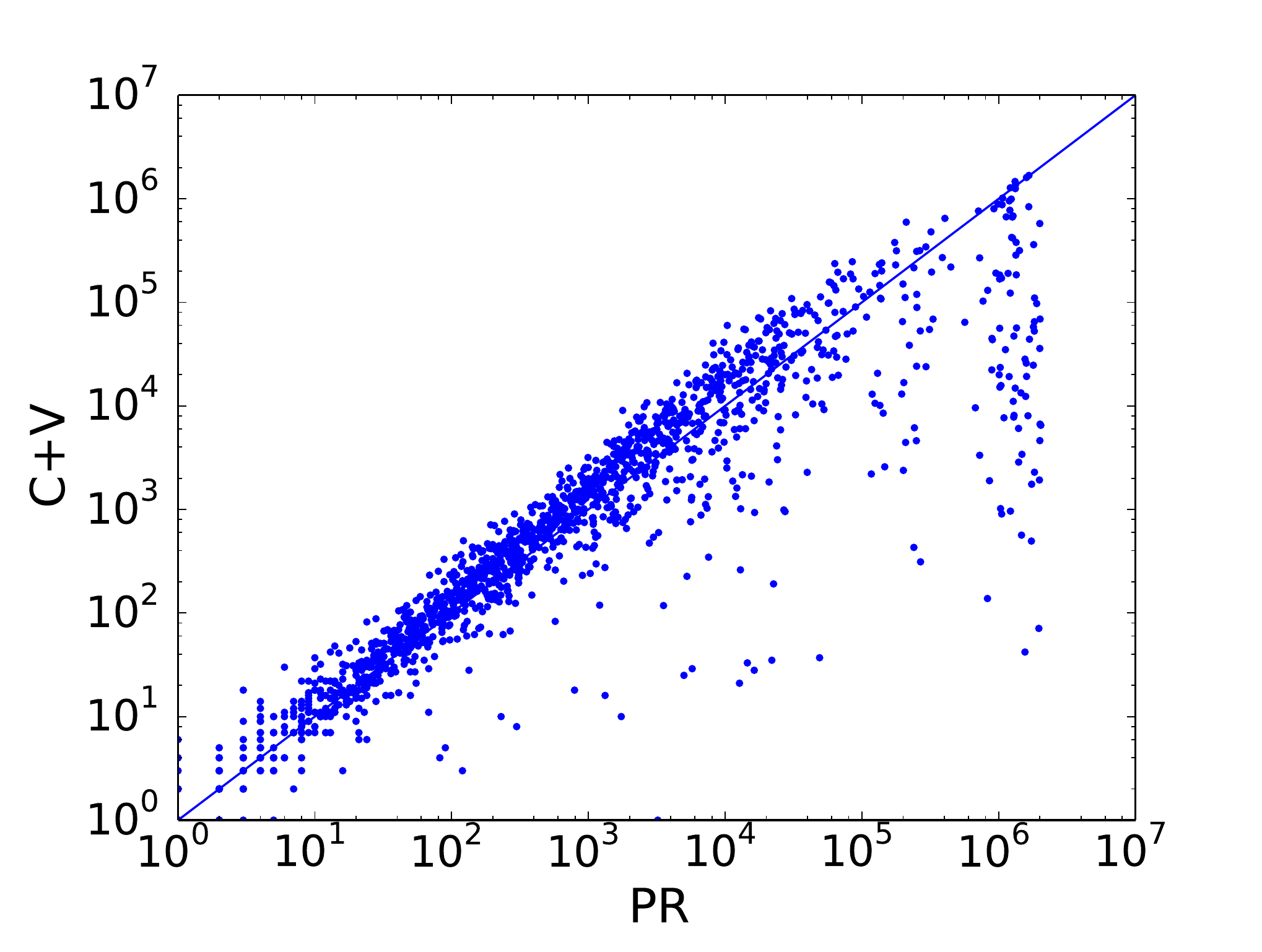}
    \caption{PR vs C+V}
  \end{subfigure}
  \hfill
  \begin{subfigure}{0.48\linewidth}
    \centering
    \includegraphics[width=\textwidth]{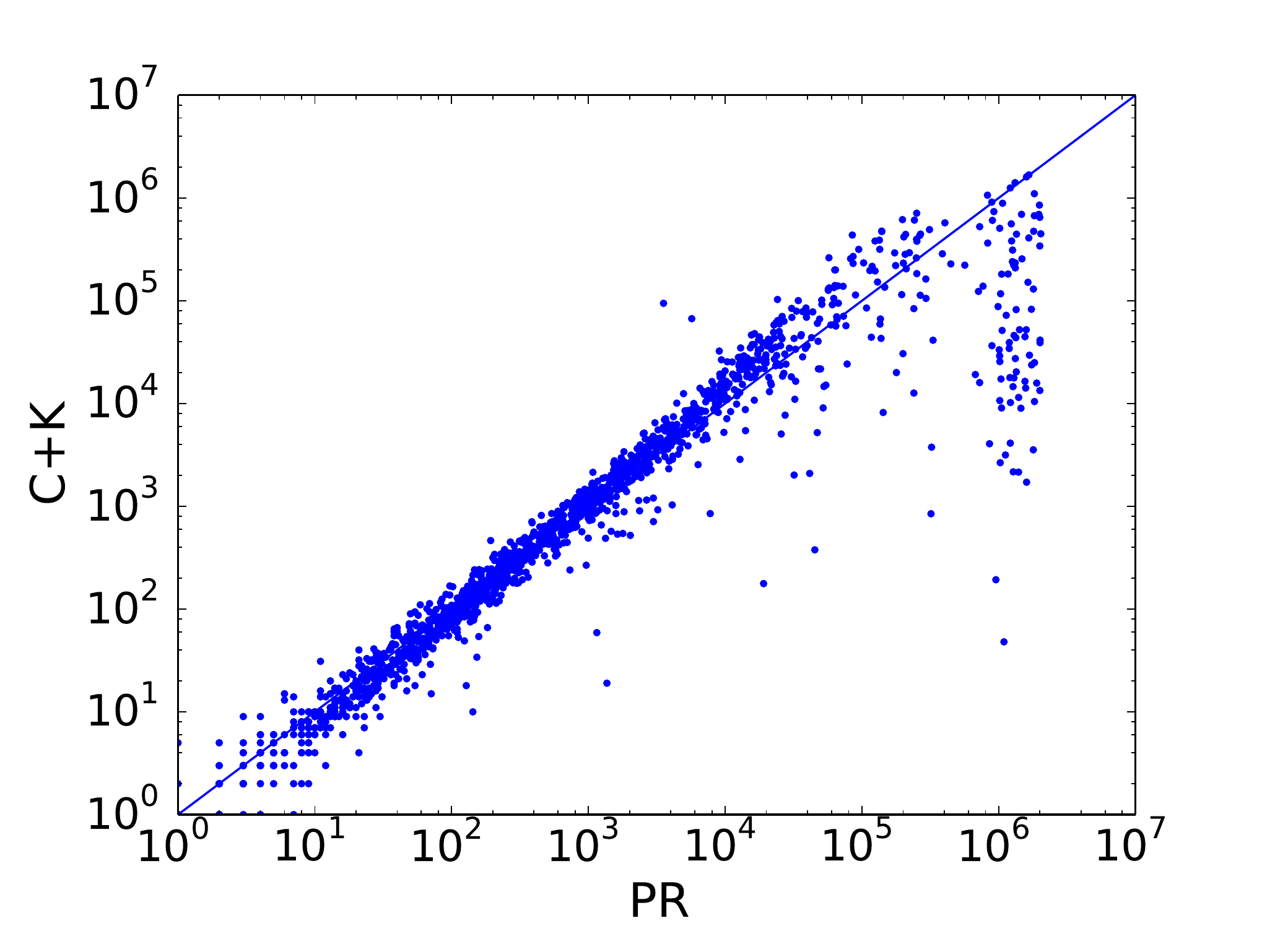}
    \caption{PR vs C+K}
  \end{subfigure}
    \hfill
  \begin{subfigure}{0.48\linewidth}
  \centering
  \includegraphics[width=\textwidth]{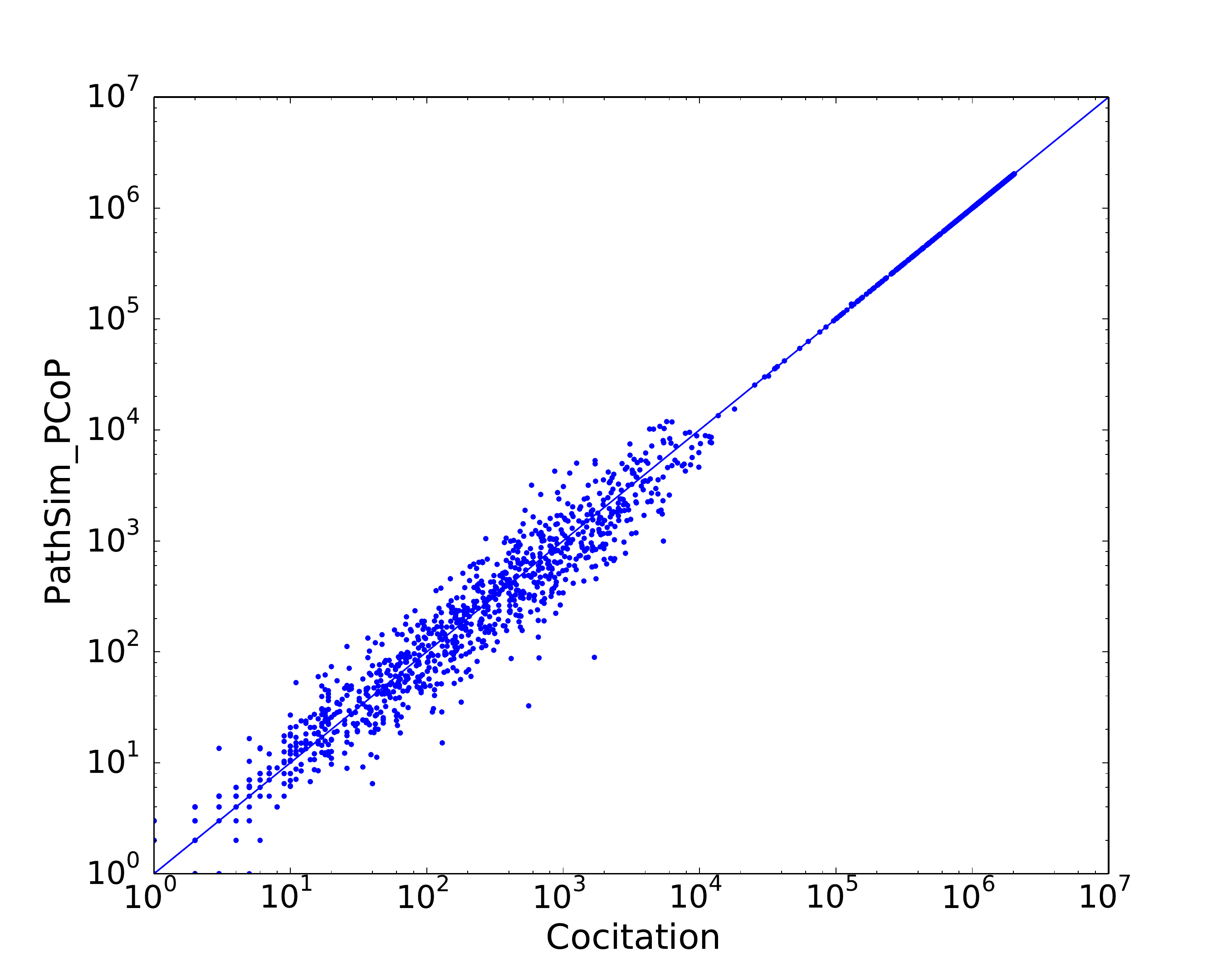}
  \caption{Cocitation vs PathSim\_PCoP}
\end{subfigure}
  \caption{Rank of hidden papers for $\delta=0$ (high correlation)}
  \label{fig:rank-correl-high}
\end{figure}

The similarity between these sets is explained by
Figure~\ref{fig:rank-correl-high} that shows a scatter plot of the
ranks of hidden papers in the different algorithms. Besides Cocitation and PathSim\_PCoP,
Collaborative Filtering and Cocitation are also highly correlated in terms of the rank of hidden
papers. This is not particularly surprising provided Collaborative Filtering and Cocitation
are using the same principles with a different weighting function. In other
words, Collaborative Filtering and Cocitation are essentially redundant algorithms.

The relations of C+X with PaperRank are somewhat different. There are
definitely a strong correlations between these methods, but some
papers see a large difference in ranks between the two methods. For
instance, two hidden papers were ranked around 1-millionth by PaperRank but
was ranked top-10 by C+A. Note also that only few hidden paper see
their rank being significantly degraded by the addition of an other
features (few papers are in the top left corner). This indicates
that the algorithms are mostly redundant, but
they are using different richer features. As such a
better way of using these features could certainly be designed.

\begin{figure}[tbp]
\begin{subfigure}{0.32\linewidth}
  \centering
  \includegraphics[width=\textwidth]{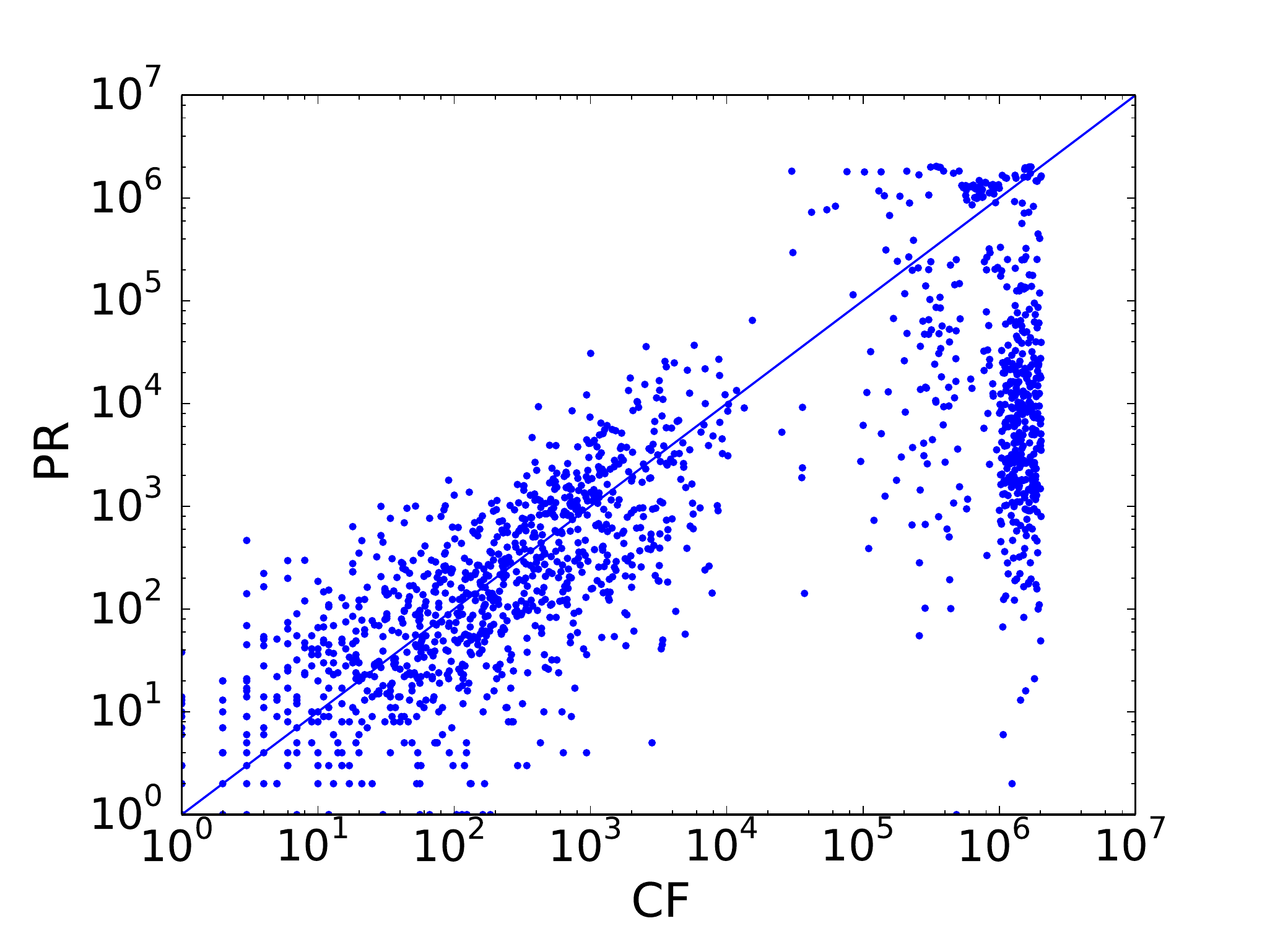}
%  \caption{CF vs RW}
\end{subfigure}
\hfill
\begin{subfigure}{0.32\linewidth}
  \centering
  \includegraphics[width=\textwidth]{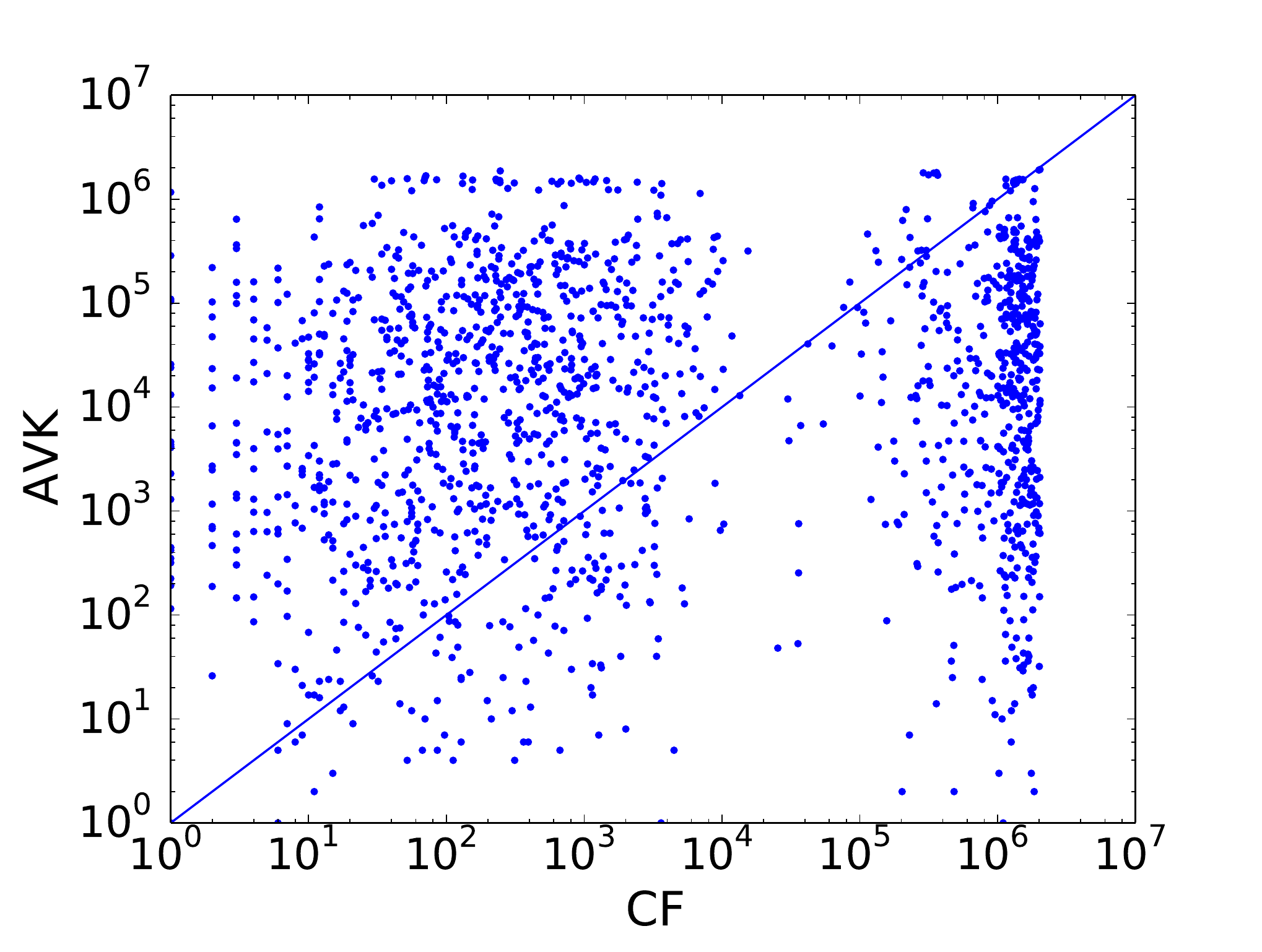}
%  \caption{CF vs AVK}
\end{subfigure}
\hfill
\begin{subfigure}{0.32\linewidth}
  \centering
  \includegraphics[width=\textwidth]{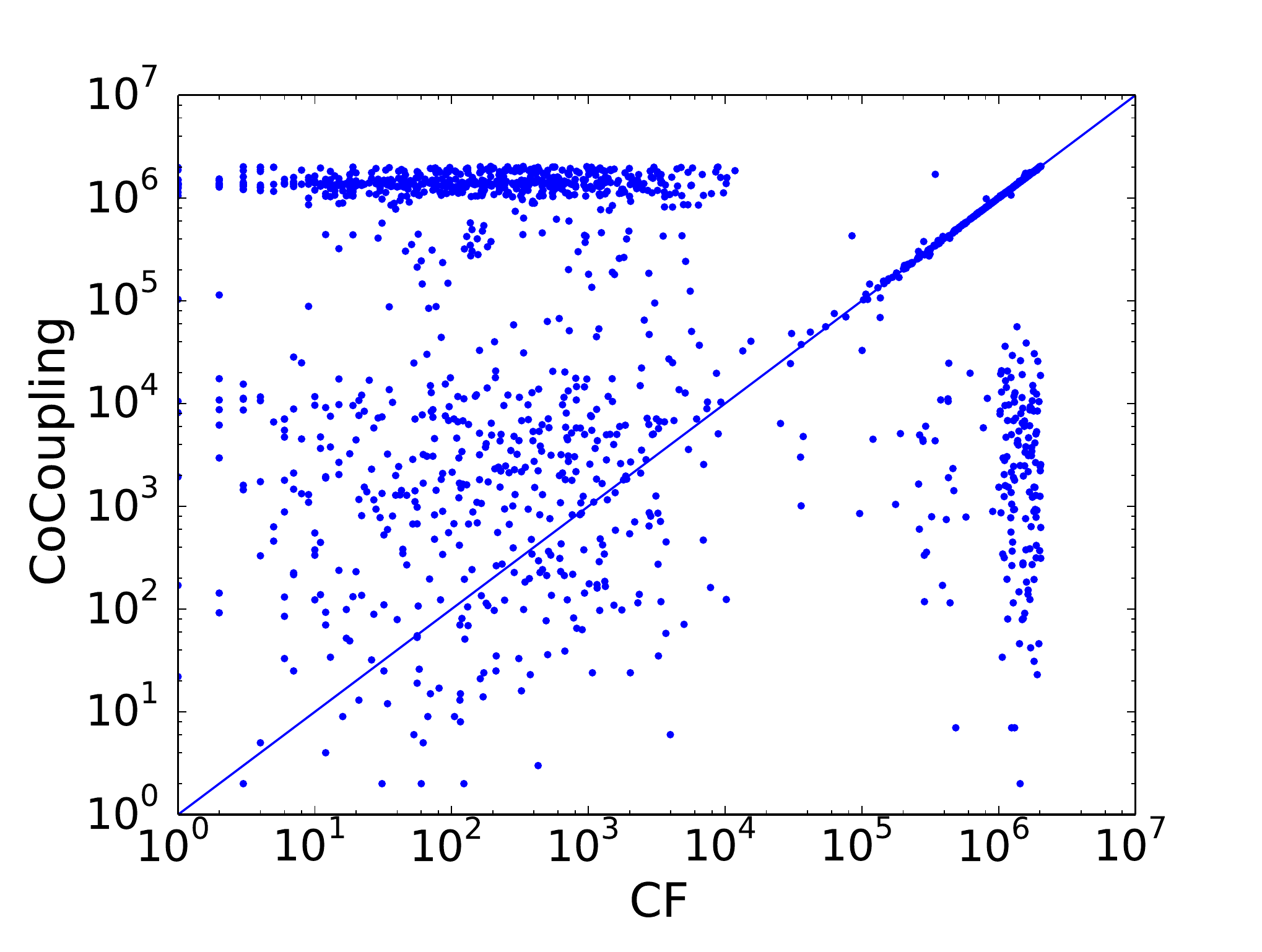}
%  \caption{CF vs Cocoupling}
\end{subfigure}

\hspace{.32\linewidth}
\hfill
\begin{subfigure}{0.32\linewidth}
  \centering
  \includegraphics[width=\textwidth]{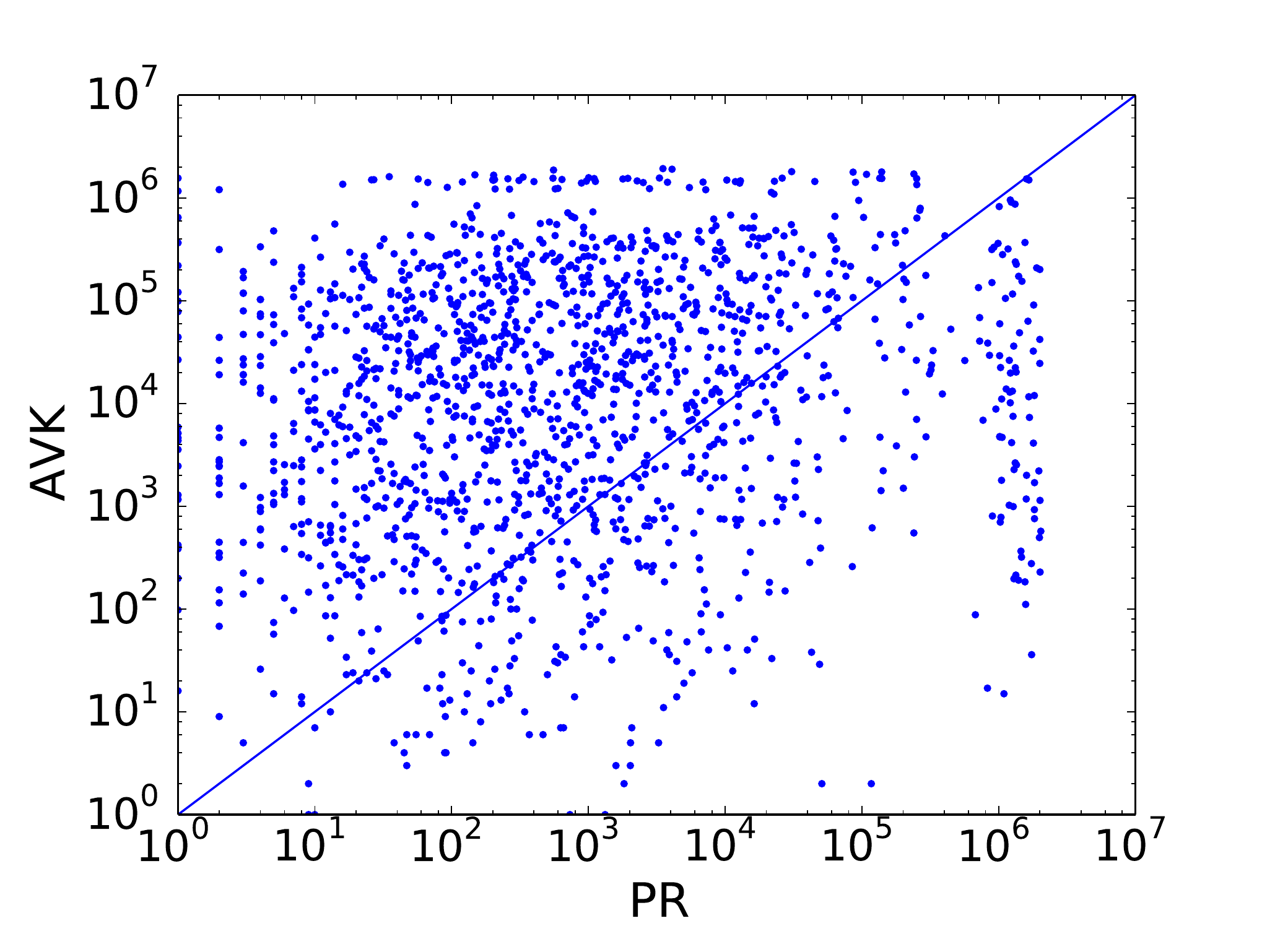}
%  \caption{RW vs AVK}
\end{subfigure}
\hfill
\begin{subfigure}{0.32\linewidth}
  \centering
  \includegraphics[width=\textwidth]{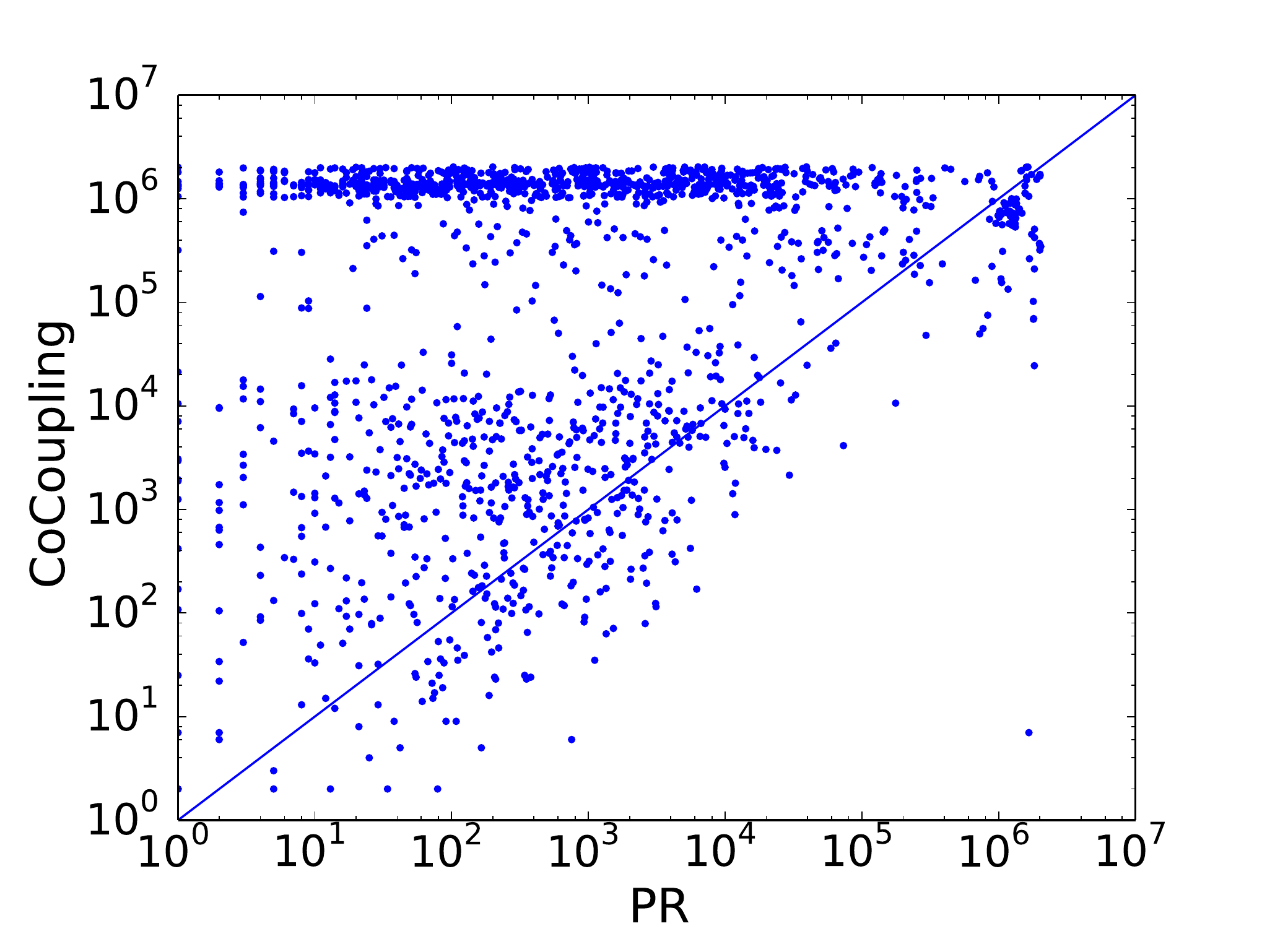}
%  \caption{RW vs CoCoupling}
\end{subfigure}

\hfill
\begin{subfigure}{0.32\linewidth}
  \centering
  \includegraphics[width=\textwidth]{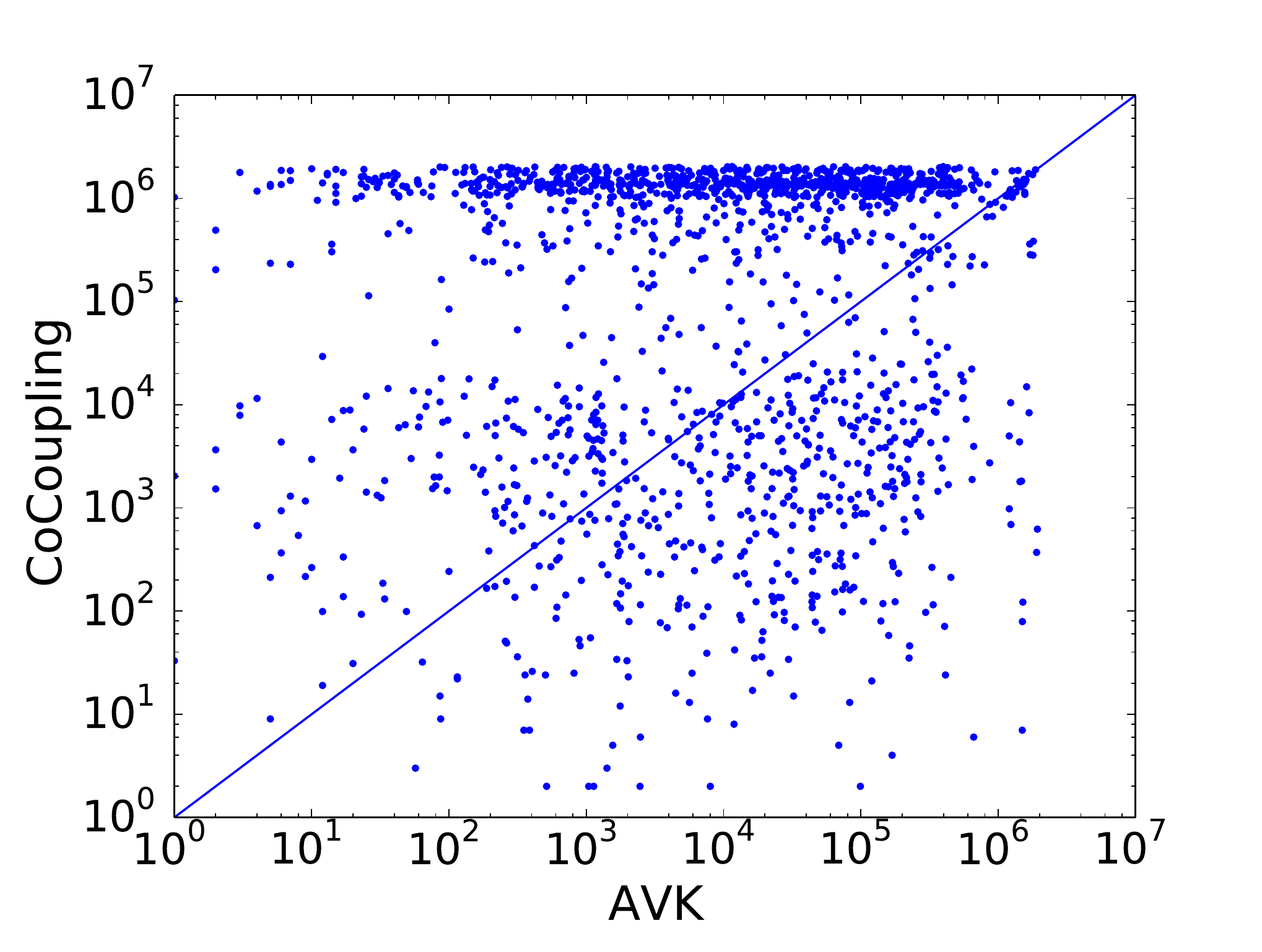}
%  \caption{AVK vs CoCoupling}
\end{subfigure}

\caption{Rank of hidden papers for $\delta=0$ (low correlation)}
\label{fig:rank-correl-low}
\vspace{-2em}
\end{figure}

Figure~\ref{fig:rank-correl-low} shows the correlation of ranks
between the remaining algorithms. C+A, C+V, C+K, Cocitation and PathSim\_PCoP are
not included because of their high correlation with either PaperRank or CF.

The rank comparison of Collaborative Filtering and Cocoupling reveals an interesting
structure. Notice that there are some hidden papers with highly
correlated with ranks over $10^5$. Digging manually in the data show
that these hidden papers are not cocited with a seed paper nor are
they cociting a common paper with a seed paper. Obviously these papers
can not be found by either method. This phenomena explains the denser
region of that scatter plots with rank over $10^5$ for Collaborative Filtering and
CoCoupling.

Collaborative Filtering and PaperRank show some correlation on the papers of rank less than
$10^4$, though the papers that are not cocited with a seed paper are
essentially randomly ordered by Collaborative Filtering.

Cocoupling does not appear to be an interesting algorithm in our
test. Indeed, Cocoupling mostly worsens the rank of hidden papers
compared to PaperRank (the hidden papers are mostly located in the upper left
region).

The logAVK method does not correlate with any other method, nor does it
seem to mostly worsen the performance of the paper nor improve them
compared to another method. logAVK does provide a completely different
perspective on the data than the other algorithms. This is not
particularly surprising since it is the only method that does not
consider the citation information.

\subsection{Peeking into the Future}

The current way of estimating the quality of a paper relies on
identifying the papers that were hidden from the list of references of
a particular paper. That experiment assumes that the author of each
paper is a data point in the ground truth. But authors are imperfect
and may not have known some papers. Rather than
using a single paper to evaluate the quality of a
recommendation, we suggest to use all the future publications.

To quantify the relevance of a recommendation, we define three metrics
 to explore different aspects of the problem.

\paragraph{Relevance-r}
For each pair of papers $<i,j>$, where $i$ is a recommended paper and $j$ is a seed paper,
we define co-cited probability as:
$$PrCo(i,j) = \frac{|C_{i,j}|}{|C_{i}|} $$
where $C_{i,j}$ denotes papers citing both $i$ and $j$ in the future and $C_{i}$ denotes
papers citing $i$ in the future.
Then,
the relevance of a recommended paper to the seed papers is:
$$Relevance(i) = \frac{\sum_{j \in S} PrCo(i,j)}{|S|} $$
Now we can evaluate the quality of a citation recommendation algorithm by the average
relevance for top K results:
$$Relevance@K = \frac{\sum_{i \in top K} Relevance(i)}{K} $$

\paragraph{Relevance-rb}
The relevance-r between a recommended paper and seed papers could
be biased by a few frequently co-cited pairs. To address this problem, we propose a binary
version of co-cited probability that just consider about whether there is a paper citing both
$i$ and $j$ in the future.
\[
PrCo(i,j) =
\begin{cases}
  1 & $$\exists C_{i,j}$$  \text{ in the future} \\
  0 & \text{otherwise}\\
\end{cases}
\]

\paragraph{Relevance-rbd} Note that we are actually interested not
only in making good recommendation, but also in making links between
papers that were not previously seen as relevant. This version of
Relevance only considers the cocitation of a seed-recommended
pairs that were not previously cocited.
\[
PrCo(i,j) =
\begin{cases}
  1 & $$\exists C_{i,j}$$  \text{ in the future and not in the past} \\
  0 & \text{otherwise}\\
\end{cases}
\]

\begin{figure}[tbp]
\centering
\begin{subfigure}{0.49\linewidth}
  \centering
  \includegraphics[width=\linewidth]{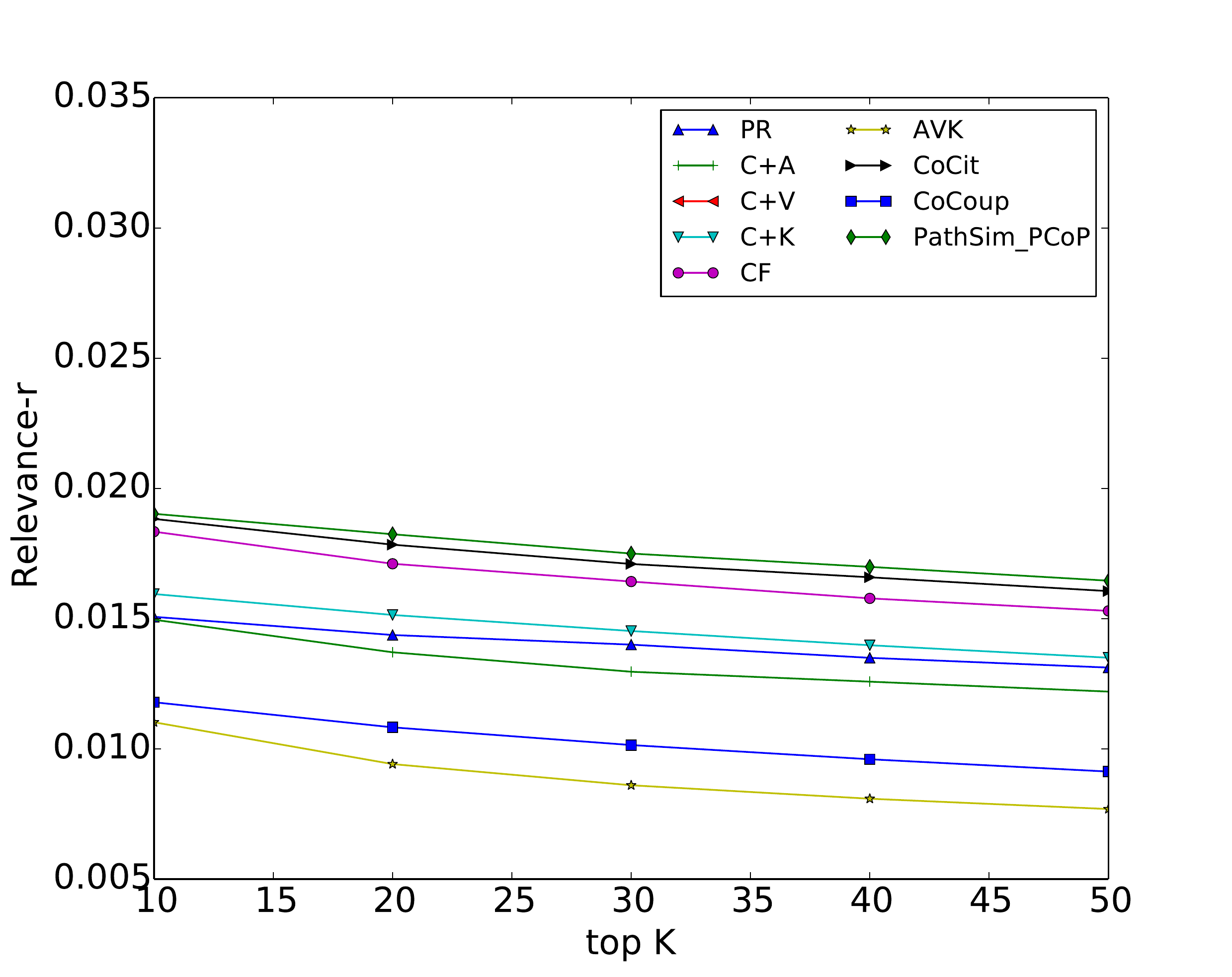}
  \caption{$\delta=0$}
\end{subfigure}
% \hfill
% \begin{subfigure}{0.3\linewidth}
%   \centering
%   \includegraphics[width=\linewidth]{plots/deg2_r.pdf}
%   \caption{$\delta \leq 2$}
% \end{subfigure}
\hfill
\begin{subfigure}{0.49\linewidth}
  \centering
  \includegraphics[width=\linewidth]{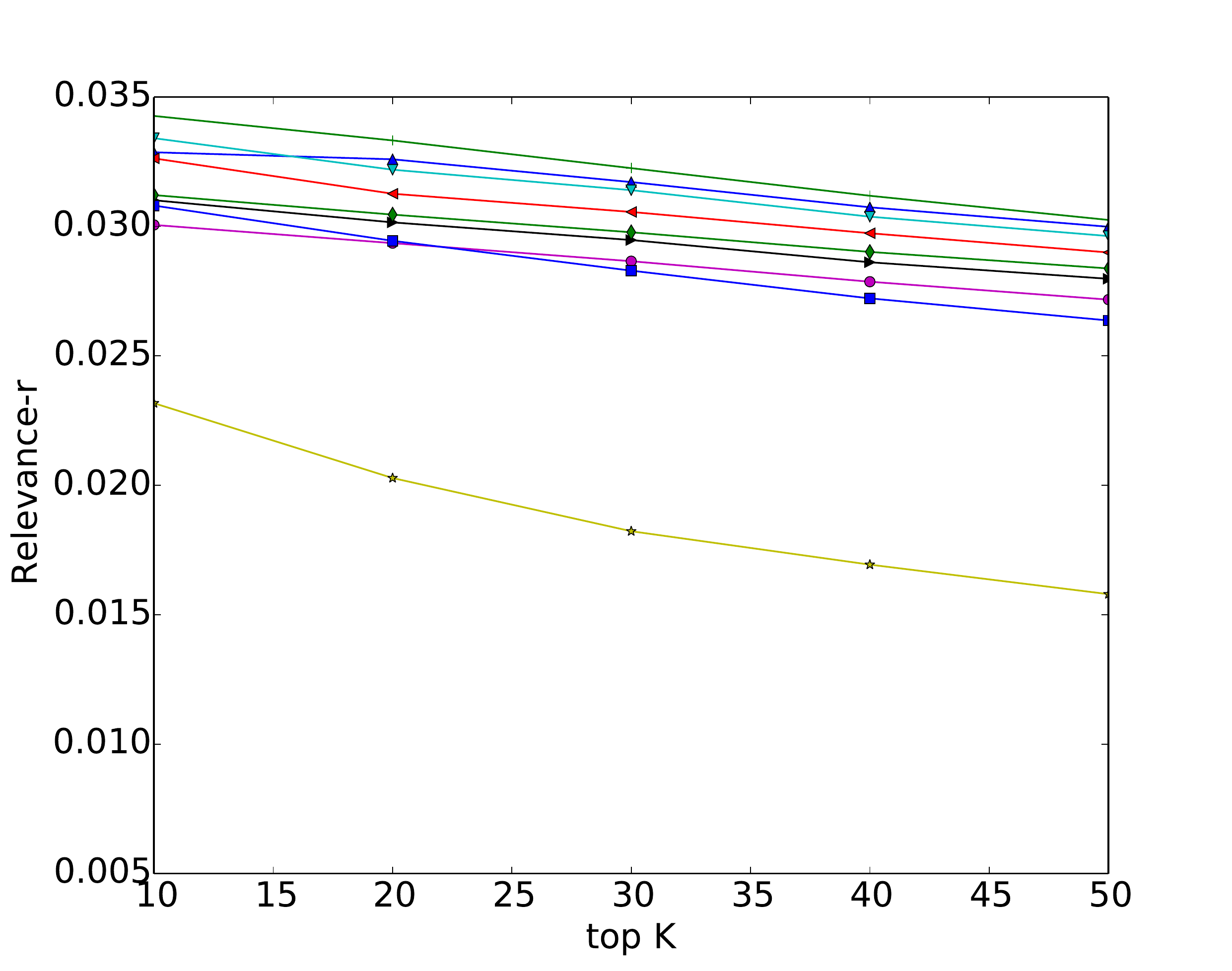}
  \caption{$\delta \leq 4$}
\end{subfigure}
\begin{subfigure}{0.49\linewidth}
  \centering
  \includegraphics[width=\linewidth]{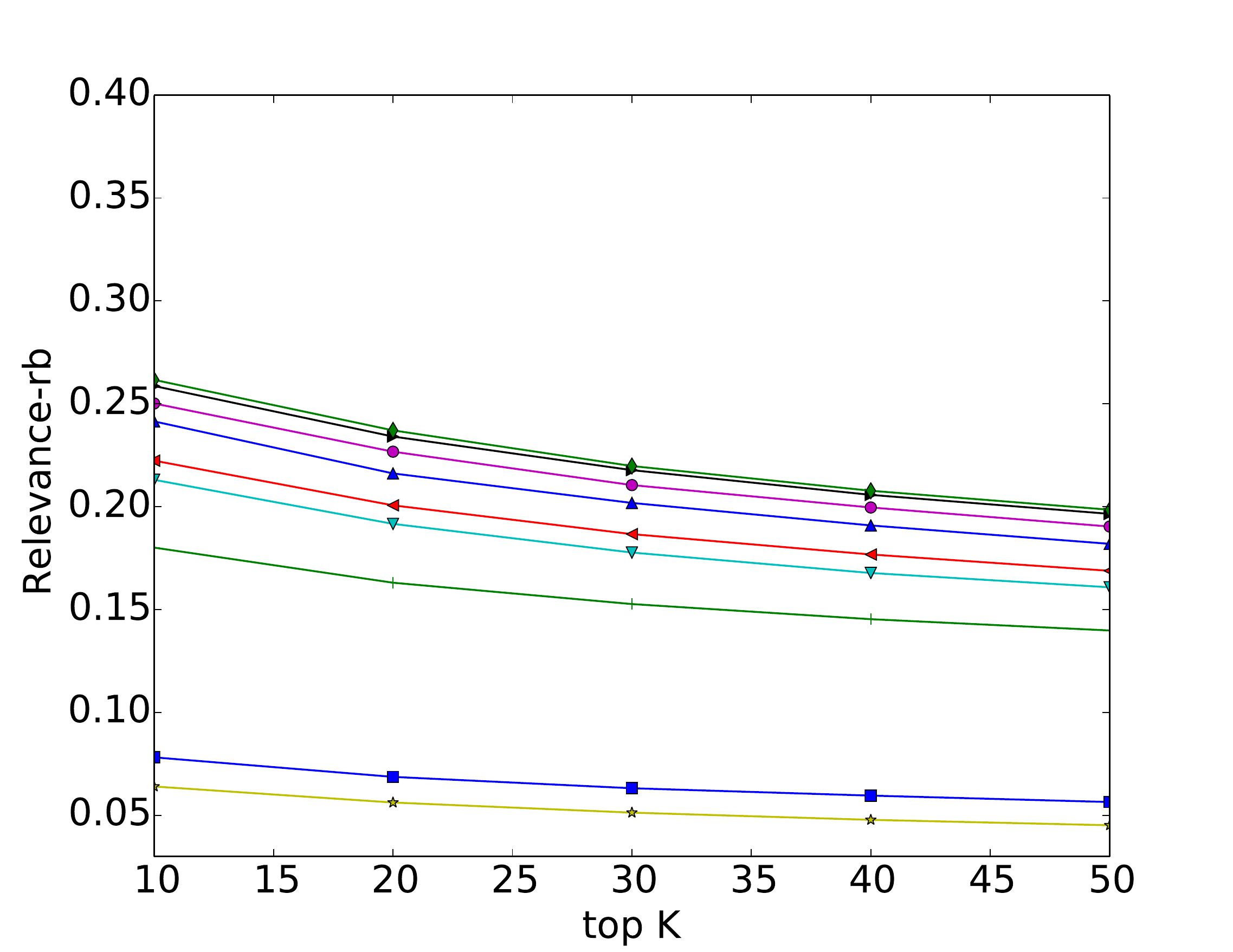}
  \caption{$\delta=0$}
\end{subfigure}
% \hfill
% \begin{subfigure}{0.3\linewidth}
%   \centering
%   \includegraphics[width=\linewidth]{plots/deg2_rb.pdf}
%   \caption{$\delta \leq 2$}
% \end{subfigure}
\hfill
\begin{subfigure}{0.49\linewidth}
  \centering
  \includegraphics[width=\linewidth]{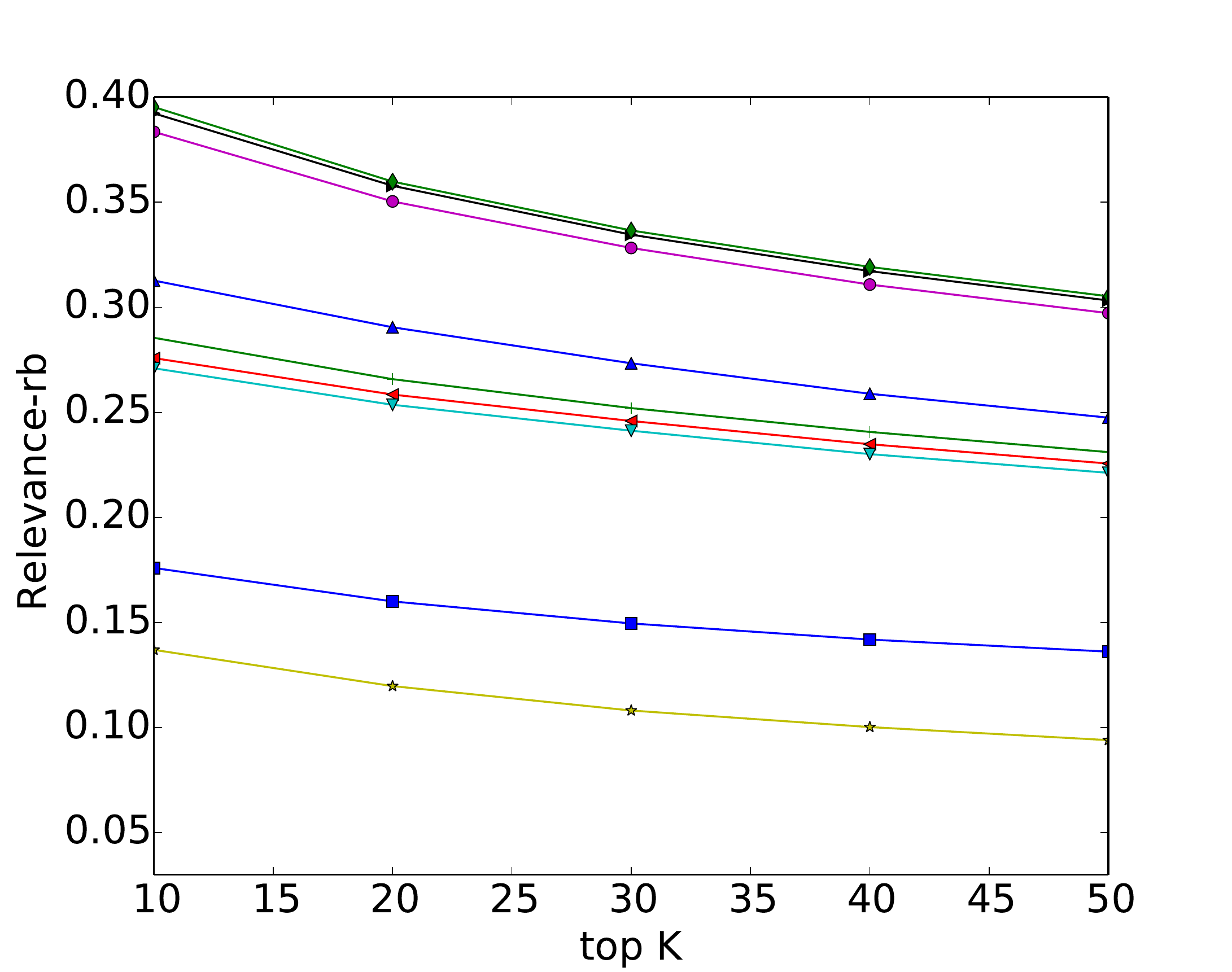}
  \caption{$\delta \leq 4$}
\end{subfigure}
\begin{subfigure}{0.49\linewidth}
  \centering
  \includegraphics[width=\linewidth]{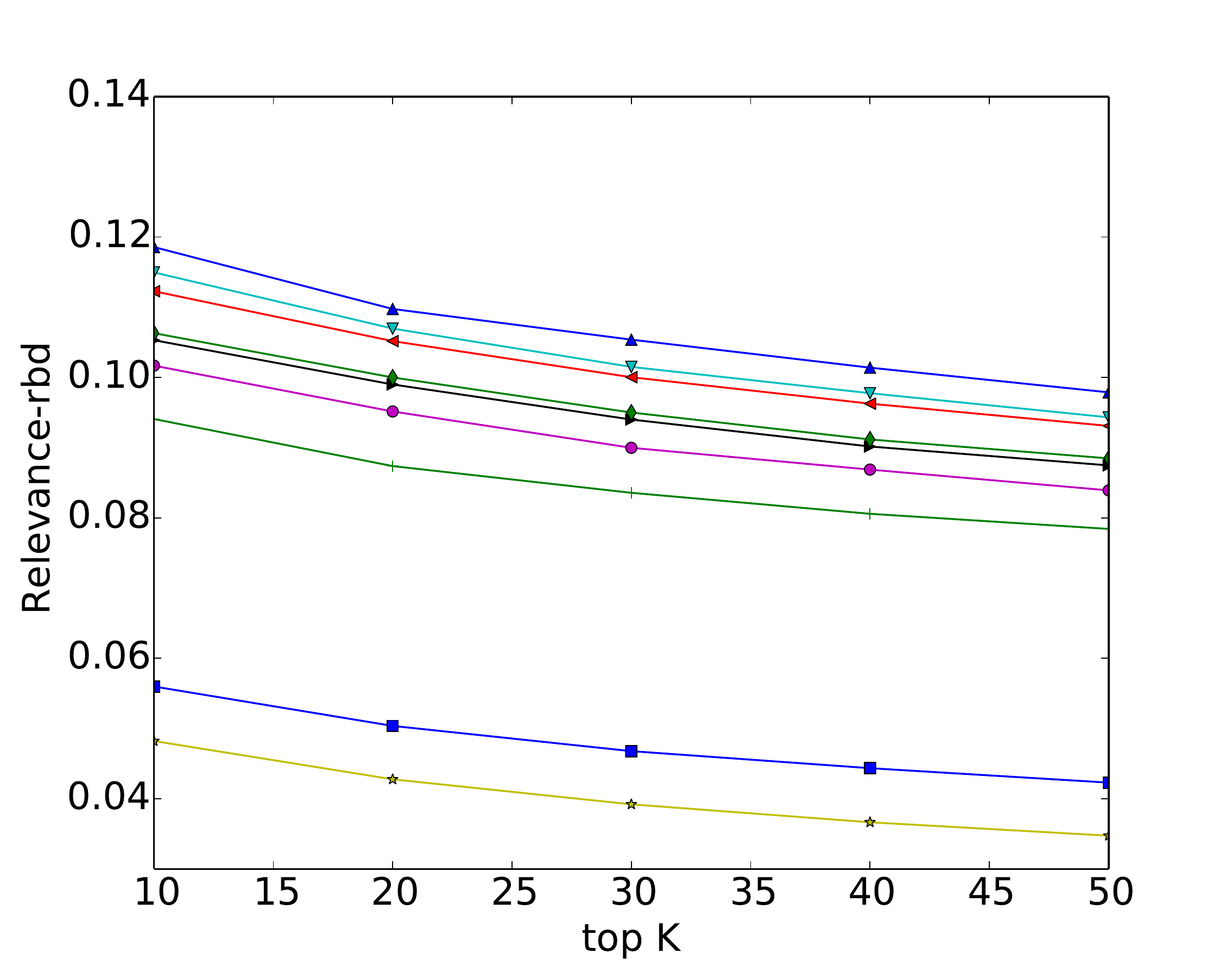}
  \caption{$\delta=0$}
\end{subfigure}
% \hfill
% \begin{subfigure}{0.3\linewidth}
%   \centering
%   \includegraphics[width=\linewidth]{plots/deg2_rbd.pdf}
%   \caption{$\delta \leq 2$}
% \end{subfigure}
\hfill
\begin{subfigure}{0.49\linewidth}
  \centering
  \includegraphics[width=\linewidth]{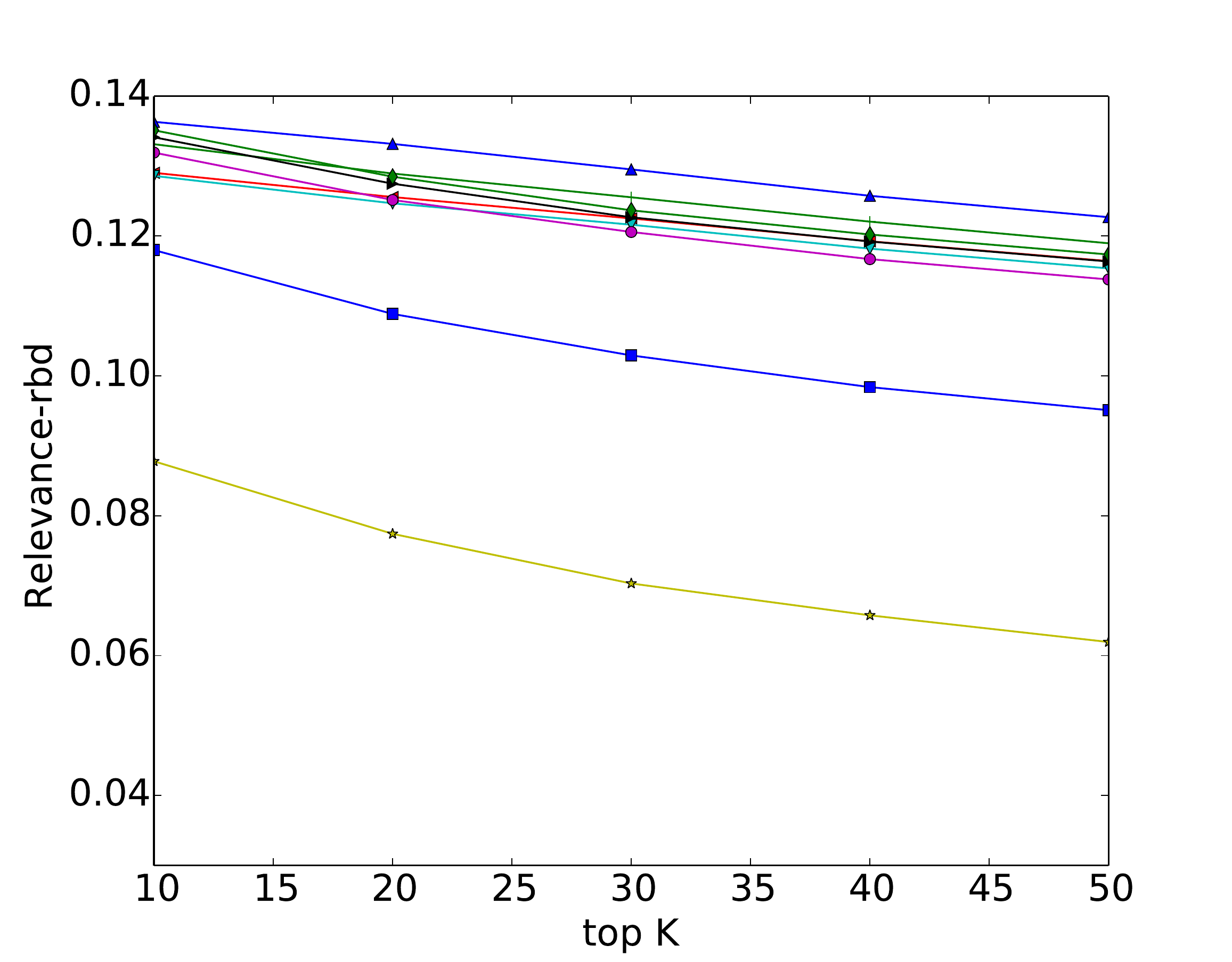}
  \caption{$\delta \leq 4$}
\end{subfigure}
\caption{Relevance}
\label{fig:relevance}
\vspace*{-3em}
\end{figure}

We computed the three relevance metrics on the same instances of the
problem we run before. We report the results of that experiment in
Figure~\ref{fig:relevance}.

Not surprisingly, the relevance decreases
when the number of returned papers increases. But the relevance does
not decrease as fast as one could expect. For instance on $\delta =
0$, the  relevance-r of algorithm C+V decreases from .013 to .011
when $k$ goes from 10 to 50. It means that 1.3\% of the future
citation to the top-10 papers recommended by C+V were in papers that
also cited a seed paper; while the relevance-r of top-50 was
1.1\%. In other words, the 50th paper recommended by C+V is not much
more irrelevant than the 10th.

Not surprisingly, current cocitations is a good predictor of future
cocitation: The Collaborative Filtering, PathSim\_PCoP and Cocitation algorithm perform usually best on
the relevance-r and relevance-rb metrics. Though when looking at
relevance-rbd that removes the citations that were already known in the
present, Collaborative Filtering, PathSim\_PCoP and Cocitation no longer are the better algorithms. PaperRank is
the algorithm that find the most relevant relations that were not known before.

It is also interesting to see that over 20\% of the recommended-seed
pairs for PaperRank will be cited in the future and half of these pairs were
not known at the time. This suggests that the algorithms we test are
actually much more helpful in practice than simple recall tests suggest.
The logAVK method also performs interestingly. About 6\% of the
recommended-seed pairs will be cited in the future (at top-10) and
most of them have not been cited before (5\% at top-10).

\begin{table}[htbp]
\caption{Upper bound for $\delta = 0$}
%\centering
\resizebox{\linewidth}{!}{
  \begin{tabular}{c|rrrrr}
 \hline
  \textbf{Metric} & \textbf{top-10} & \textbf{top-20} & \textbf{top-30} & \textbf{top-40}& \textbf{top-50}\\ \hline
  \rmfamily Relevance\textunderscore r & \rmfamily 0.286093  & \rmfamily 0.205920 & \rmfamily 0.170241  & \rmfamily 0.148972 & \rmfamily 0.134334\\
  \rmfamily Relevance\textunderscore rb & \rmfamily 0.880969  & \rmfamily 0.702677 & \rmfamily 0.590164  & \rmfamily 0.520564 & \rmfamily 0.473333\\
  \rmfamily Relevance\textunderscore rbd & \rmfamily 0.778027  & \rmfamily 0.605512 & \rmfamily 0.505168  & \rmfamily 0.443790 & \rmfamily 0.402499\\
  \hline
\end{tabular}
}
\label{tab:upper0}
\end{table}

\begin{table}[htbp]
\vspace{-2em}
\caption{Upper bound for $\delta \leq 4$}
%\centering
\resizebox{\linewidth}{!}{
\begin{tabular}{c|rrrrr}
 \hline
  \textbf{Metric} & \textbf{top-10} & \textbf{top-20} & \textbf{top-30} & \textbf{top-40}& \textbf{top-50}\\ \hline
  \rmfamily Relevance\textunderscore r & \rmfamily 0.368085  & \rmfamily 0.272176 & \rmfamily 0.225938  & \rmfamily 0.197591 & \rmfamily 0.178001\\
  \rmfamily Relevance\textunderscore rb & \rmfamily 0.998309  & \rmfamily 0.975889 & \rmfamily 0.879426  & \rmfamily 0.787065 & \rmfamily 0.717206\\
  \rmfamily Relevance\textunderscore rbd & \rmfamily 0.868585  & \rmfamily 0.768658 & \rmfamily 0.674373  & \rmfamily 0.600422 & \rmfamily 0.545786\\
  \hline
\end{tabular}
}
\label{tab:upper4}
\end{table}

We computed upper bounds on the relevance metrics to quantify how good
the different algorithms are. Indeed, we can use the knowledge of the
future to easily compute for each query the relevance of each paper
and greedily pick the $k$ papers of highest relevance. We report the
upper bound on best relevance for $\delta=0$ in Table~\ref{tab:upper0}
and for $\delta \leq 4$ in Table~\ref{tab:upper4}.  The upper bounds
are much higher than the relevance of the algorithms: a factor of 10
on relevance-r, 4 on relevance-rb, and 5 on relevance-rbd. This
indicates that there is a significant room for improvement in our
paper recommendation tasks: there are better set of papers that will
be cocited with the seed papers than the methods are recommending.
% \begin{table}[tbp]
% \caption{Relevance for seed papers}
% \centering
% \begin{tabular}{crrrrr}
%  \hline
%   \textbf{Metric} & \textbf{Recall@10} & \textbf{Recall@20} & \textbf{Recall@30} & \textbf{Recall@40}& \textbf{Recall@50}\\
%   \rmfamily Relenvance\textunderscore r & \rmfamily 0.339207  & \rmfamily 0.196567 & \rmfamily 0.0436877  & \rmfamily 0.0164212 & \rmfamily 0.0076858\\
%   \rmfamily Relenvance\textunderscore rb & \rmfamily 1  & \rmfamily 0.869316 & \rmfamily 0.184351  & \rmfamily 0.067185 & \rmfamily 0.0211347\\
%   \rmfamily Relenvance\textunderscore rbd & \rmfamily 0.89171  & \rmfamily 0.681205 & \rmfamily 0.149816  & \rmfamily 0.0586109 & \rmfamily 0.017401\\
%   \hline
% \end{tabular}
% \label{tab:seed}
% \end{table}

\subsection{Implications for a practical system?}

We evaluated many algorithms, namely PaperRank, Collaborative Filtering,
Cocitation, Cocoupling, C+A, C+V, C+K, PathSim\_PCoP and LogAVK. The evaluation was
performed across different tests, metrics, and by looking at different
slices of the solution space. We present here a summary of the
discussion with a focus on selecting algorithms for inclusion in a
practical system.

Cocitation, PathSim\_PCoP and Collaborative Filtering are variations of the same
algorithm and their performance are hard to distinguish. (See
correlation in Figure~\ref{fig:rank-correl-high} and the difference in
recommendation in Table~\ref{tab:diff-delta2}). There is no point in including both
algorithms in a system: we will pick
Collaborative Filtering.

Cocoupling is often one of the worst algorithm and is essentially
worse than PaperRank. (See correlation plot in
Figure~\ref{fig:rank-correl-low}). As such, we do not believe it makes
sense to include Cocoupling if any variants of PaperRank were to be included.

The C+V, C+A, C+K algorithms are somewhat correlated to PaperRank but
they exhibit improvement for many cases (see Figure~\ref{fig:rank-correl-high}). C+K
has the highest recall on the $\delta=0$ study case (see
Figure~\ref{fig:recall-low-degree}), and C+A and C+K showed the highest relevance-r in the
$\delta \leq 4$ case (see Figure~\ref{fig:relevance}). We believe one
of these methods should be included in practice, but more
work in integrating metadata in the recommendation is necessary.\looseness=-1

The logAVK algorithm provides a much lower recall than the other
algorithm (See Figure~\ref{fig:recall-low-degree} for example). However, we believe it
could be of some interest to discover loosely connected
papers. Indeed, it returns papers that are very different from the
other methods (See Table~\ref{tab:diff-delta2}) while having a relevance that is
within a factor of 2 or 3 of the other algorithms (see Figure~\ref{fig:relevance}
for $\delta=0$). We believe that LogAVK could provide a view of the
problem that is complementary to the one provided by the citation
based methods.

\subsection{Fast C+X Recommendation}

For a practical citation recommendation system, the efficiency
of underlying recommendation algorithm is also important.
The running time of random walk based methods typically
depends on the size of input graph and thus tends to be more expensive. While
some other method like collaborative filtering essentially computes
the weighted co-citation relationships and thus does not need to take the global
graph into account. Previous work~\cite{Jia18} has shown that LocRank, which is a local
version of PaperRank, is as effective as PaperRank while being much faster than
PaperRank and CF. Here we will explore the local methods for C+X.

We define a local induced subgraph of a query $q$: $G_q=(V_q,E_q)$, where $V_q$ contains all nodes
in $S$ and any node which is a neighbor of at least one seed paper:
$$V_q = S \cup S_n$$
where $S_n$ denotes
$$S_n =  \bigcup_{s \in S} \{ v:v \in Adj(s)\} $$
$E_q$ remains all citation relationships between nodes in $V_q$. In other words, $G_q$ is
the subgraph induced by the distance 1 neighborhood of the seed papers.
Then, we extend $G_q$ to a heterogeneous graph $\mathcal{G}_q$ by adding metadata information of $V_q$, local C+X computes a random walk on $\mathcal{G}_q$.

\begin{table}[tbp]
\caption{Performance for fast recommendation}
\centering
\begin{tabular}{c|c|rrr}
 \hline
  \textbf{Method} & \textbf{Sec/query} & \textbf{Recall@10} & \textbf{Recall@20}& \textbf{Recall@50} \\\hline
  \rmfamily C+A  & \rmfamily 3.82 & \rmfamily 0.230617  & \rmfamily 0.318206 & \rmfamily 0.463204\\
  \rmfamily C+V  & \rmfamily 3.59 & \rmfamily 0.230531  & \rmfamily 0.323125 & \rmfamily 0.461898\\
  \rmfamily C+K  & \rmfamily 3.91 & \rmfamily 0.231308  & \rmfamily 0.315485 & \rmfamily 0.461507\\
  \rmfamily C+A\_Local  & \rmfamily 0.24 & \rmfamily 0.229565  & \rmfamily 0.308260 & \rmfamily 0.448141  \\
  \rmfamily C+V\_Local  & \rmfamily 0.22 & \rmfamily 0.215549  & \rmfamily 0.296508 & \rmfamily 0.436924   \\
  \rmfamily C+K\_Local  & \rmfamily 0.25 & \rmfamily 0.221331 & \rmfamily 0.303953 & \rmfamily 0.446463  \\
  \hline
\end{tabular}
\label{tab:fast}
\end{table}

\begin{figure}[tbp]
  \begin{center}
    \includegraphics[width=.7\linewidth]{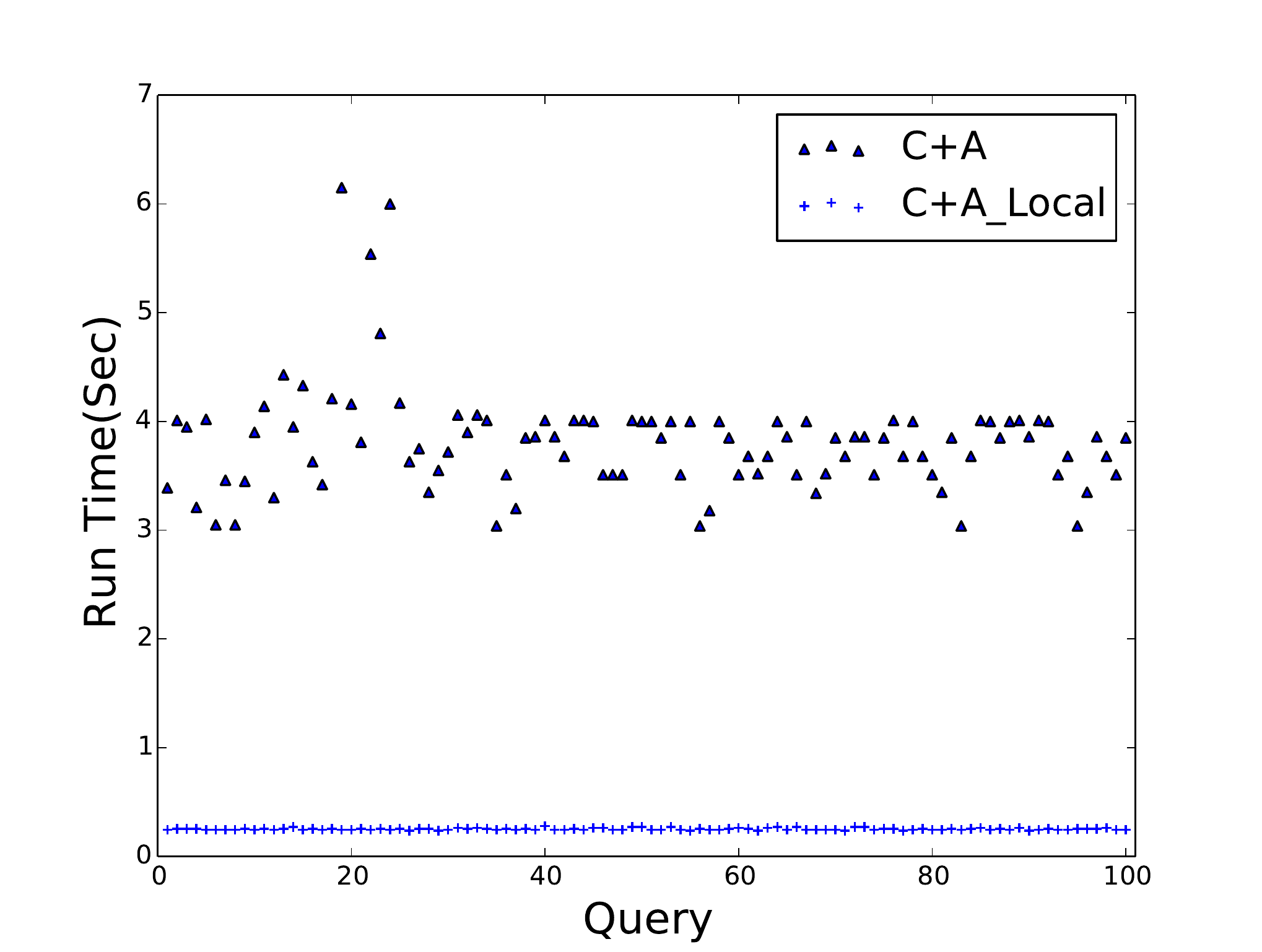}
  \end{center}
  \caption{Runtime on 100 instance queries}
  \label{fig:runtime}
  \vspace{-2em}
\end{figure}

In our experiments, all codes are written in C++ and the graphs are represented in Compressed
Row Storage format for compact storage. The codes are compiled with g++
4.8.2 with option -O3. The codes are run on 1 core of an Intel(R) Xeon CPU
E-5-2623 @ 3.00GHz processor.

As we can see from the Table~\ref{tab:fast}, the column Sec/query shows the average
runtime per query for each method. In general, local C+X is 15x faster than original
methods. It is not surprising because the runtime of local methods only depends on the size of local
induced graph, while original ones are global ranking methods; Moreover, a local induced
graph tends to have a smaller diameter, which means local C+X can reach the convergence
within less iterations. In Figure~\ref{fig:runtime}, we take C+A as example and show the
runtime for 100 randomly sampled independent queries. Note that since we remove the query paper q and all papers published
after q from the citation graph to simulate the time when the query paper was
being written, the size of the global graph is different for queries with different
publication date.

Besides the much better efficiency of local C+X methods, the quality of recommendation is
still competitive comparing with original methods. Essentially, local methods
are tradeoffs between the upper bound of recall and the efficiency. It turns out that
they have equivalent abilities to find hidden papers as global methods, which demonstrates
that many findable hidden papers are actually neighbors of seed papers.

\section{Conclusion}
\label{sec:ccl}

This manuscript investigates the problem of recommending a set of papers to
extend a query composed of a set of known paper. This problem is
common in academic recommender systems and academic search engines.
The two most common citation recommendation algorithms, PaperRank
and Collaborative Filtering, do not uniformly discover relevant papers;
they mostly find a set of papers that are highly connected to the
query by citations. Unfortunately, real-world citation patterns are
not as obvious to find since about 50\% cocited papers do not have a
direct connection. The key to improving the quality of an academic
recommender system lies in identifying those loosely connected, yet
relevant, papers. While we consider the problem of identifying highly
connected papers essentially solved by the existing methods.

We provided two ways of discovering citations that use the metadata of
the papers rather than their citation patterns, LogAVK that only uses
the metadata and the C+X algorithm which combine the citation pattern
and the metadata. The C+K and C+A algorithms are promising in
retrieving papers that are loosely connected to the query. Despite
logAVK is about 3 times less relevant than PaperRank, it identifies
papers that are known to be important and which are likely to be
unknown to the user and the community.

Using a single test/metric to qualify algorithms provides
an incomplete picture of how good an algorithm is. We
believe that the proposed relevance metrics provide additional
insights in the quality and desirability of algorithms.

Future works will focus on building new models for integrating
metadata inside a random walk framework to connect better similar
papers that are not connected by citations. Currently existing methods
require to return a large number of papers to achieve desired
recall. Therefore, there is a need in presenting a set of paper to a
user in a structured non-list format so that the user can easily
navigate the recommendation and identify the papers that appear more
relevant.

\section*{Acknowledgment}

This material is based upon work supported by the National Science Foundation under Grant No. 1652442.

%%%%%%%%%%%%%%%%%%%%%%%%%%%%%%%%%%%%%%%%%%%%%%%%%%%%%%%%%%%%%%%%%%%%%%%%%

\bibliographystyle{spbasic}
\bibliography{paper}

\end{document}